\documentclass[11pt]{article}% добавляется twoside для двусторонней печати

\usepackage[T1]{fontenc}
\usepackage[cp1251]{inputenc}
\usepackage{textcomp}
\usepackage[centertags]{amsmath}
\usepackage[mediummath]{nccmath}
\usepackage{amsfonts}
\usepackage{amssymb}
\usepackage{enumitem}
\usepackage[hypertex]{hyperref}
\usepackage{graphicx}
\usepackage{graphbox}
\usepackage[numbers,sort&compress]{natbib}% упорядочивает ссылки

\usepackage{paperinitial}% Установка параметров страницы

\paperinitialization{15mm}{15mm}{15mm}{15mm}{2pt}{10pt}

\DeclareMathOperator{\pr}{pr}

\DeclareMathOperator{\Sp}{Sp}

\newcommand{\ttg}{\stepcounter{equation}\tag{\theequation}}

\newcommand{\lan}{\langle}
\newcommand{\ran}{\rangle}
\newcommand{\bs}{\boldsymbol}

\newcommand{\e}{\varepsilon}
\newcommand{\vf}{\varphi}

\newcommand{\s}{\sigma}

\newcommand{\al}{\alpha}
\newcommand{\be}{\beta}
\newcommand{\ga}{\gamma}

\newcommand{\de}{\delta}
\newcommand{\De}{\Delta}

\newcommand{\la}{\lambda}

\newcommand{\spx}{\mathbf{x}}

\newcommand{\spp}{\mathbf{p}}
\newcommand{\spk}{\mathbf{k}}

\newcommand{\R}{\mathbb{R}}

\begin{document}
\allowdisplaybreaks[4]% позволяет переносить многострочные формулы
\frenchspacing% уменьшение пробелов после запятых
%\setlength{\unitlength}{1pt}% устанавливает единицу длины в окружении picture
%\selectlanguage{english}

\title{{\Large\textbf{Coherent radiation of photons by particle wave packets}}}

\date{}

\author{P.O. Kazinski\thanks{E-mail: \texttt{kpo@phys.tsu.ru}}\;\, and T.V. Solovyev\thanks{E-mail: \texttt{nevybrown@mail.ru}}\\[0.5em]
{\normalsize Physics Faculty, Tomsk State University, Tomsk 634050, Russia}
}

\maketitle

\begin{abstract}

The radiation of photons by electrons is investigated in the framework of quantum electrodynamics up to the second order in the coupling constant $e$. The $N$-particle, coherent, and thermal initial states are considered and the forms of the electron wave packets are taken into account. The explicit expressions for the intensity of radiation and the inclusive probability to record a photon are obtained. It is found that there are three processes in this order of perturbation theory where the electron wave packet radiates coherently and can be regarded as a charged fluid even on integrating over the final states of the electron, i.e., in considering the inclusive probabilities and intensity of radiation. These processes are stimulated radiation by an electron, coherent radiation from a beam of particles, and reradiation of a photon in the Compton process. We obtain the explicit expressions for the intensity of radiation and the inclusive probability to record a photon for these processes. As particular cases, we consider: stimulated transition radiation produced by an electron wave packet traversing a mirror and backlighted by a laser wave, reradiation of photons in a coherent state by an electron wave packet. In the latter case, we deduce that the wave packet of a single electron can be endowed with the susceptibility tensor and this tensor has the same form as for an electron plasma in the small recoil limit.

\end{abstract}

\section{Introduction}

There is a long-lived and actively debated problem regarding the radiation of photons by electron wave packets. A naive interpretation of this process inspired by the original Schr\"{o}dinger interpretation of the wave function suggests that the electron wave packets should radiate as a charged fluid on average \cite{IvKarl13prl,IvKarl13pra,KonPotPol,PanGov18,GovPan18,KarlZhev19,PanGov19,PanGov21,Talebi16}. On the other hand, there are the direct calculations in the framework of quantum electrodynamics (QED) and the experiments showing that this is not the case \cite{MarcuseI,MarcuseII,SundMil90,LapFedEber93,PMHK08,CorsPeat11,WCCP16,AngDiPiaz18,Remez19,KdGdAR21,Wong21}, at least, for certain processes when the so-called inclusive radiation probabilities are considered, i.e., only the quantum numbers of radiated photons are measured, whereas the escaping electrons and other particles are unobserved. The present-day consensus (see, e.g., \cite{Wong21,PupKarl22}) consists in that there are a few QED processes where the inclusive radiation probability can be obtained by considering the electron wave packet as a charged fluid. As a rule, in most radiation processes a great amount of information about the profile of the wave packet of a radiating particle is lost in the inclusive radiation probability. In the present paper, we systematically investigate the inclusive radiation probabilities in QED up to the second order in the coupling constant $e$ and list all such processes where the electron wave packet radiates coherently as some kind of a charged fluid on average.

It was argued in \cite{MarcuseII} that coherent radiation from a beam of charged particles strongly depends on the forms of wave packets of particles constituting the beam. The explicit formulas for such coherent radiation showing that it is determined by the Dirac currents of particle wave functions were derived in \cite{pra103} in the leading order of perturbation theory. The technical reason for appearance of coherent radiation from particle wave packets is the presence of through lines in the Feynman diagrams for the expansion of the $S$-matrix. On squaring the transition amplitude, these diagrams give rise to interference terms depending on the diagonal of the transition currents which, in turn, result in contributions to inclusive probabilities corresponding to coherent radiation from particle wave packets. It turns out that apart from coherent radiation from a beam of charged particles there are two other processes in the second order perturbation theory where coherent radiation from an electron wave packet can be observed. These processes are stimulated radiation created by an electron wave packet \cite{PanGov18,GovPan18,PanGov19,PanGov21,Talebi16} and reradiation of photons by an electron wave packet in the Compton scattering. As for the latter process, due to the fact that the electron wave packet can be regarded as a charged fluid in this process, the electron wave packet possesses an electric susceptibility. The susceptibility tensor coincides with the susceptibility tensor of an electron plasma in the small recoil limit despite the fact that only one electron participates in the process.

In considering scattering of particles as plane waves, in particular, in evaluating the cross-section, the diagrams with through lines and, more generally, the disconnected contributions to the $S$-matrix are excluded \cite{WeinbergB.12}. The disconnected contributions to the $S$-matrix vanish almost everywhere in the momentum space and so, for scattering of particles with wave functions possessing definite momenta, they can be omitted. In particular, by definition, only the connected part of the $S$-matrix contributes to the invariant amplitude $M$ for plane waves since otherwise $|M|^2$ appearing in the cross-section is not defined. However, the disconnected contributions to the $S$-matrix become relevant when scattering of wave packets of a general form is studied. In this case, these contributions are of the same order of magnitude as those coming from the connected part of the $S$-matrix. Moreover, such terms change the familiar coupling constant orders of contributions of the processes to the observed probabilities. For example, for scattering of a plane-wave photon by a plane-wave electron in a vacuum (the Compton process), the leading contribution of the perturbation theory to the cross-section is of the order $e^4$. Notwithstanding, as we shall show, the leading nontrivial contribution to the inclusive probability to record a photon scattered by an electron wave packet with nontrivial structure is of the order $e^2$. In particular, these reasonings imply that the general formulas presented, for example, in Sec. 4.1 of \cite{KoSeSch92}, in Sec. 2 of \cite{Karl17}, and in Sec. 3 of \cite{Ivan22}, where the $S$-matrix for scattering of wave packets of a general form is given in terms of the plane-wave invariant amplitude $M$, have to be refined. 

%The contribution of through lines of the Feynman diagrams gives rise to the terms in the inclusive probability where the wave functions of some particles participating in process evolve freely and so do not experience a quantum recoil. This is the reason why a coherent emission is possible even in considering inclusive probabilities.

The paper is organized as follows. In Sec. \ref{GenForm}, we start with a formal statement of the scattering problem and list all the contributions to the inclusive radiation probability in QED up to the second order in $e$. We specify the initial states and the projector realizing the measurement in the final state. As the initial states of photons, we consider the coherent and thermal states. As for the electron initial states, we take the $N$-particle Fock state and the thermal one. Then evaluation of the inclusive probability is reduced to evaluation of the corresponding traces, which is readily performed in the Bargmann-Fock representation. Thus we obtain the general formulas for all the second order contributions to the inclusive probability to record a photon and to the radiation intensity. In particular, we correct the formulas of \cite{BKL5}, where among other things the radiation of photons at finite temperature was studied, and introduce the general formula for the effective polarization tensor. The nonperturbative approach to take into account the photons in the initial coherent state with large amplitude -- the so-called Furry picture -- is also considered in this section. Section \ref{Stim_Rad} is devoted to a detailed description of stimulated radiation produced by electron wave packets. For definiteness, we consider stimulated transition radiation from a Dirac particle wave packet crossing an ideally conducting plate irradiated by a laser wave. Note that stimulated transition radiation was also studied in \cite{ZarLomNer80,IvanLomon81} where the profiles of particle wave packets and the disconnected contributions to the $S$-matrix were not taken into account. In this section, we derive the explicit expression for the radiation intensity and show that this process can be employed for imaging the profile of an electron wave function by using the developed methods of noninvasive diagnostics of beams of charged particles. We also deduce the selection rules for this process when the one-particle density matrix possesses certain symmetries. In Sec. \ref{Electr_Suscept}, we investigate reradiation of photons by electrons. We establish that in this process, in the given order of perturbation theory, the electron wave packet can be regarded as a charged fluid and find the susceptibility tensor of this fluid. The symmetries of the one-particle density matrix give rise to the selection rules for reradiated photons as in the case of a dispersive medium with the same symmetries. In Conclusion we summarize the results.

We adopt the following notation. We denote by indices $A$, $B$, $\ldots$ the space-time index, $\mu$, and the point of the space-time $x$. The Greek indices $\al$, $\be$, $\bar{\al}$, $\bar{\be}$, $\ldots$ denote the quantum numbers of particle states. The summation (integration) over repeated indices is always understood unless otherwise stated. We also suppose that the quantum states of particles are normalized to unity in some sufficiently large volume $V$. As for the expansion of quantum fields in terms of the mode functions, we use the agreements chosen in the paper \cite{pra103}. In order to conform the notations of \cite{pra103} and \cite{ippccb}, we use interchangeably the star and the bar as the sign of complex conjugation. The bar over the Dirac spinor means as usual the Dirac conjugate. Furthermore, wherever it does not lead to misunderstanding, we use the matrix notation. For example,
\begin{equation}
    \bar{a}a\equiv a^*a\equiv a^*_\al a_\al,\qquad \bar{d}Dd\equiv d^*Dd\equiv d^*_\al D_{\al\bar{\al}}d_{\bar{\al}},\qquad \text{etc.}
\end{equation}
We use the system of units such that $\hbar=c=1$ and $e^2=4\pi\al$, where $\al$ is the fine structure constant. The Minkowski metric is taken with the mostly minus signature.

\section{Inclusive probability}\label{GenForm}

In order to find the intensity of radiation and the inclusive probability to record a photon, it is convenient to employ the Bargmann-Fock representation \cite{BerezMSQ1.4}. A brief synopsis of the main relations of this formalism can be found in Appendix A of the paper \cite{ippccb}. For the reader convenience, we collect the most frequently used formulas in Appendix \ref{BF_represnt_Ap_A}. Some details of the bulky calculations arising in evaluating inclusive probabilities are removed to Appendix \ref{Traces_Ap_B}.

\subsection{Initial and final states}

As the initial states, we will take
\begin{enumerate}[wide, labelwidth=!, labelindent=0pt]
  \item The $N$-particle Fock state of fermions (electrons)
\begin{equation}\label{N_part_state}
\begin{split}
    |\Phi\ran&=k  \vf^1_{\al_1}\cdots \vf^N_{\al_N} a^\dag_{\al_1}\cdots a^\dag_{\al_N}|0\ran,\\
    \Phi(\bar{a})&=k(\vf^1\bar{a})\cdots (\vf^N \bar{a}),  \qquad|k|^{-2}=\det(\bar{\vf}^i\vf^j),
\end{split}
\end{equation}
where $i=\overline{1,N}$, $j=\overline{1,N}$.
  \item The normalized coherent state
\begin{equation}\label{coher_state}
    \Phi(\bar{c})=e^{(\bar{c}-\bar{d})d},
\end{equation}
where $\bar{d}d$ is the average number of particles in this state.
  \item The density matrix of a general form
\begin{equation}
    \lan\bar{a}|R|a\ran=\frac{1}{\sqrt{N!M!}}\rho_{\al_N\cdots\al_1|\bar{\al}_1\cdots\bar{\al}_M} \bar{a}_{\al_1}\cdots\bar{a}_{\al_N} a_{\bar{\al}_N}\cdots a_{\bar{\al}_1},\qquad \rho_{\al_N\cdots\al_1|\al_1\cdots \al_N}=1,
\end{equation}
where the summation over $N$ and $M$ is understood.
  \item The thermal state of noninteracting particles
\begin{equation}
    \lan\bar{a}|R|a\ran=\exp(\bar{a}e^{-\be\tilde{\e}}a)/Z,\qquad \ln Z=-\epsilon\Sp\ln(1-\epsilon e^{-\be\tilde{\e}}),
\end{equation}
where $\tilde\e_{\al\bar{\al}}:=\e_{\al\bar{\al}}-\mu q_{\al\bar{\al}}$ and $\e_{\al\bar{\al}}$, $q_{\al\bar{\al}}$ are the one-particle operators of energy and charge, $[\e,q]=0$, and $\mu$ is the chemical potential, $\epsilon=1$ for Bose-Einstein statistics and $\epsilon=-1$ for Fermi-Dirac statistics.
\end{enumerate}
Notice that, in the cases $1$, $4$, the density matrix commutes with the particle number operator. The density matrix is defined in the interaction picture at the instant of time $t=0$. At some initial instant of time $t=t_{in}$, it becomes
\begin{equation}
    \lan\bar{a}|R|a\ran= \frac{1}{\sqrt{N!M!}}\rho_{\al_N\cdots\al_1|\bar{\al}_1\cdots\bar{\al}_M} \bar{a}_{\al_1}\cdots\bar{a}_{\al_N} a_{\bar{\al}_N}\cdots a_{\bar{\al}_1}e^{-i(\e_{\al_1}+\cdot \e_{\al_N} -\e_{\bar{\al}_1}-\cdots \e_{\bar{\al}_N}) t_{in}},
\end{equation}
where it is assumed that the one-particle states are the stationary ones.

Let us introduce the one-particle density matrix,
\begin{equation}\label{rho_1}
    \rho^{(N,1)}_{\al\bar{\al}}:=\rho_{\al_N\cdots\al_2\al|\bar{\al}\al_2\cdots\al_N},\qquad\Sp(Ra^\dag_{\bar{\al}} a_\al)=N\rho^{(N,1)}_{\al\bar{\al}},
\end{equation}
and two-particle density matrix,
\begin{equation}\label{rho_2}
    \rho^{(N,2)}_{\al_2\al_1|\bar{\al}_1\bar{\al}_2}=\rho_{\al_N\cdots\al_3\al_2\al_1|\bar{\al}_1\bar{\al}_2\al_3\cdots\al_N}, \qquad\Sp(Ra^\dag_{\bar{\al}_1}a^\dag_{\bar{\al}_2} a_{\al_2}a_{\al_1})=N(N-1)\rho^{(N,2)}_{\al_2\al_1|\bar{\al}_1\bar{\al}_2}.
\end{equation}
The summation over $N$ is understood in the second formulas in \eqref{rho_1} and \eqref{rho_2}.

In the particular case of the $N$-particle state of fermions, we have
\begin{equation}\label{rho_1_2_N}
\begin{split}
    \rho^{(N,1)}_{\al\bar{\al}}&=\frac{|k|^2}{N}\sum_{k=1}^N\det
    \left[
      \begin{array}{ccccc}
        \bar{\vf}^1\vf^1 & \cdots & \bar{\vf}^1_{\bar{\al}}\vf^k_{\al} & \cdots & \bar{\vf}^1\vf^N \\
        \vdots &   & \vdots &   & \vdots \\
        \bar{\vf}^N\vf^1 & \cdots & \bar{\vf}^N_{\bar{\al}}\vf^k_{\al} & \cdots & \bar{\vf}^N\vf^N \\
      \end{array}
    \right],\\
    \rho^{(N,2)}_{\al_2\al_1|\bar{\al}_1\bar{\al}_2}&=\frac{|k|^2}{N(N-1)}\bigg\{\underset{l<k}{\sum_{k,l=1}^N} \det
    \left[
      \begin{array}{ccccccc}
        \bar{\vf}^1\vf^1 & \cdots & \bar{\vf}^1_{\bar{\al}_1}\vf^l_{\al_1} & \cdots & \bar{\vf}^1_{\bar{\al}_2}\vf^k_{\al_2} & \cdots & \bar{\vf}^1\vf^N \\
        \vdots &   & \vdots &   & \vdots &  & \vdots\\
        \bar{\vf}^N\vf^1 & \cdots & \bar{\vf}^N_{\bar{\al}_1}\vf^l_{\al_1} & \cdots & \bar{\vf}^N_{\bar{\al}_2}\vf^k_{\al_2} & \cdots & \bar{\vf}^N\vf^N \\
      \end{array}
    \right]-(\al_1\leftrightarrow\al_2)\bigg\}.
\end{split}
\end{equation}
If
\begin{equation}\label{orthogonality}
    \bar{\vf}^i\vf^j\approx\de^{ij},\qquad i,j=\overline{1,N},
\end{equation}
then $|k|^2\approx1$ and
\begin{equation}
    \rho^{(N,1)}_{\al\bar{\al}}\approx\frac{1}{N}\sum_{k=1}^N \vf_{\al}^k\bar{\vf}^k_{\bar{\al}},\qquad \rho^{(N,2)}_{\al_2\al_1|\bar{\al}_1\bar{\al}_2}=\frac{1}{N(N-1)}\sideset{}{'}\sum_{k,l=1}^N\vf^k_{\al_2}\vf^l_{\al_1} \bar{\vf}^l_{[\bar{\al}_1}\bar{\vf}^k_{\bar{\al}_2]},
\end{equation}
where the prime at the sum sign means that the term with $k=l$ is excluded and the square brackets at the pair of indices denote antisymmetrization without the factor $1/2$. In particular, the condition \eqref{orthogonality} is satisfied when the common phases of the wave functions $\vf^i_\al$ are random, uncorrelated, and equiprobably distributed, i.e.,
\begin{equation}
    \vf^i_\al=e^{i\xi_i}\tilde{\vf}^i_\al,\qquad\lan\xi_i\xi_j\ran=0,\quad i\neq j,\qquad\lan e^{i\xi_i}\ran=0,
\end{equation}
where $\xi_i$ are stochastic quantities, and the final expression for the probability to record a photon is averaged over the random phases $\xi_i$.

As for the thermal state of noninteracting fermions, we obtain
\begin{equation}\label{rho_therm_ferm}
    N\rho^{(N,1)}_{\al\bar{\al}}=(n_f)_{\al\bar{\al}},\qquad N(N-1)\rho^{(N,2)}_{\al_2\al_1\bar{\al}_2\bar{\al}_2} =(n_f)_{\al_1[\bar{\al}_1} (n_f)_{\al_2\bar{\al}_2]},
\end{equation}
where $(n_f)_{\al\bar{\al}}:=(e^{\be\tilde{\e}}+1)^{-1}_{\al\bar{\al}}$. Similarly, for bosons we have
\begin{equation}\label{rho_therm_bos}
    N\rho^{(N,1)}_{\al\bar{\al}}=(n_b)_{\al\bar{\al}},\qquad N(N-1)\rho^{(N,2)}_{\al_2\al_1\bar{\al}_2\bar{\al}_2} =(n_b)_{\al_1(\bar{\al}_1} (n_b)_{\al_2\bar{\al}_2)},
\end{equation}
where $(n_b)_{\al\bar{\al}}:=(e^{\be\tilde{\e}}-1)^{-1}_{\al\bar{\al}}$ and the parenthesis at the pair of indices mean symmetrization without the factor $1/2$.

The projector to the states in the Fock space containing at least one particle in the states singled out by the projector $D_{\al\bar{\al}}$ in the one-particle Hilbert space takes the form
\begin{equation}\label{P_D}
    P=1-:\exp(-c^\dag Dc): \, =c^\dag Dc+\cdots,\qquad D^\dag=D.
\end{equation}
Below, we will not use the property $D^2=D$. Besides, we denote $\tilde{D}:=1-D$. Notice that the operator $P$ commutes with the particle number operator.

We will consider the processes such that the initial state contains photons and electrons, whereas, in the final state, only photons are recorded or the intensity of radiation of photons with given energy and other quantum numbers is detected. Thus
\begin{equation}\label{P_D1}
    P=P_{ph}\otimes1_{e} \otimes1_{e^+},
\end{equation}
where $P_{ph}$ has the form \eqref{P_D} with $c_\ga$ being the annihilation operator of a photon with quantum numbers $\ga$. We also suppose that either $D$ is diagonal in the energy basis or the operator $D$ is taken at the instant of time $t=0$ and so, for $t=t_{out}$, it becomes
\begin{equation}
    D_{\al\bar{\al}}(out)=D_{\al\bar{\al}} e^{-i(k_{0\al}-k_{0\bar{\al}})t_{out}}.
\end{equation}
It follows from expression \eqref{P_D} that the intensity of radiation, $I_D$, in the mode with quantum numbers $\ga$ is a linear part of the inclusive probability to record a photon, $P_D$, with respect to the operator $D$, where $D_{\al\bar{\al}}=k_{0\ga}\pr^\ga_{\al\bar{\al}}$ (no summation over $\ga$) and $\pr^\ga_{\al\bar{\al}}$ is the projector to the state with quantum numbers $\ga$. Henceforth, we shall use this property in order to obtain $I_D$ from $P_D$. Besides, we will take into account only the leading terms in the coupling constant and assume that the vacuum is stable. Furthermore, we will suppose that the electrons and the photons are uncorrelated in the initial state, viz.,
\begin{equation}\label{dens_matr_ini}
    R=R_{ph}\otimes R_{e}\otimes|0\ran_{e^+}\lan0|_{e^+},
\end{equation}
where $R_{ph}$ is the density matrix of electrons and $R_e$ is the density matrix of photons.

\subsection{Perturbative treatment of photons in the initial state}
\subsubsection{Contributions to the $S$-matrix}

The radiation of a photon by an electron in the external field or in the presence of the dispersive medium is described by the operator,
\begin{equation}\label{V_oper}
    V=V^{\bar{\ga}}_{\bar{\al}\al} a^\dag_{\bar{\al}} a_{\al} c^\dag_{\bar{\ga}} -V^{\dag\ga}_{\bar{\al}\al}a^\dag_{\bar{\al}} a_{\al} c_\ga,\qquad V^{\bar{\ga}}_{\bar{\al}\al}:=-i\lan\bar{\al},e^-|j^A|\al,e^-\ran \bar{e}_A^{\bar{\ga}},
\end{equation}
of the first and higher orders in the coupling constant $e$. Here $j^A$ is the current density operator and $e^\ga_A$ are the mode functions of the photon quantum field. For example, in a vacuum, these functions can be chosen in the form of plane waves,
\begin{equation}
    e^\ga_A=\frac{e_\mu^{(\la)}(\spk)}{\sqrt{2k_0 V}}e^{-ik_\nu x^\nu},
\end{equation}
where $e_\mu^{(\la)}(\spk)$ are the polarization vectors and $\ga=(\la,\spk)$, $\la=1,2$.

Apart from the operator \eqref{V_oper}, there exists the operator
\begin{equation}
    E=E^{\ga}_{\bar{\al}\bar{\be}}a^\dag_{\bar{\al}} b^\dag_{\bar{\be}} c_{\ga},\qquad E^{\ga}_{\bar{\al}\bar{\be}}:=-i\lan\bar{\be},e^+;\bar{\al},e^-|j^A|0\ran e_A^{\ga},
\end{equation}
where $b^\dag_\be$ is the positron creation operator, of the first order in the coupling constant. However, this operator gives the second order contribution for the processes we consider.

The Compton process is specified by the operators,
\begin{equation}
\begin{aligned}
    B&=B_1+B_2+B_3,&&\\
    B_1&=B^{\bar{\ga}\ga}_{1\bar{\al}\al}a^\dag_{\bar{\al}} a_{\al} c^\dag_{\bar{\ga}} c_{\ga},&\qquad B^{\bar{\ga}\ga}_{1\bar{\al}\al}&=2 B^{AB}_{\bar{\al}\al}\bar{e}_A^{\bar{\ga}} e_B^\ga,\\
    B_2&=B^{\ga_1\ga_2}_{2\bar{\al}\al}a^\dag_{\bar{\al}} a_{\al} c_{\ga_1} c_{\ga_2},&\qquad B^{\ga_1\ga_2}_{2\bar{\al}\al}&= B^{AB}_{\bar{\al}\al}e_A^{\ga_1} e_B^{\ga_2},\\
    B_3&=B^{\bar{\ga}_1\bar{\ga}_2}_{3\bar{\al}\al}a^\dag_{\bar{\al}} a_{\al} c^\dag_{\bar{\ga}_1} c^\dag_{\bar{\ga}_2},&\qquad B^{\bar{\ga}_1\bar{\ga}_2}_{3\bar{\al}\al}&= B^{AB}_{\bar{\al}\al}\bar{e}_A^{\bar{\ga}_1} \bar{e}_B^{\bar{\ga}_2},\\
\end{aligned}
\end{equation}
of the second order in the coupling constant. Here we have introduced the notation
\begin{equation}
    B^{AB}_{\bar{\al}\al}=-\frac12 \lan\bar{\al},e^-|:j^A j^B:|\al,e^-\ran.
\end{equation}
Of the same order are the contributions of the mass operator,
\begin{equation}
    M=M_{\bar{\al}\al}a^\dag_{\bar{\al}} a_{\al},
\end{equation}
and the polarization operator,
\begin{equation}
\begin{aligned}
    \Pi&=\Pi_1+\Pi_2+\Pi_3,\\
    \Pi_1&=\Pi^{\bar{\ga}\ga}_{1} c^\dag_{\bar{\ga}} c_{\ga},&\qquad \Pi^{\bar{\ga}\ga}_{1}&=2 \Pi^{AB}\bar{e}_A^{\bar{\ga}} e_B^\ga,\\
    \Pi_2&=\Pi^{\ga_1\ga_2}_{2} c_{\ga_1} c_{\ga_2},&\qquad \Pi^{\ga_1\ga_2}_{2}&=\Pi^{AB} e_A^{\ga_1} e_B^{\ga_2},\\
    \Pi_3&=\Pi^{\bar{\ga}_1\bar{\ga}_2}_{3} c^\dag_{\bar{\ga}_1} c^\dag_{\bar{\ga}_2},&\qquad \Pi^{\bar{\ga}_1\bar{\ga}_2}_{3}&=\Pi^{AB} \bar{e}_A^{\bar{\ga}_1} \bar{e}_B^{\bar{\ga}_2},\\
\end{aligned}
\end{equation}
where
\begin{equation}
    \Pi^{AB}=-\frac12 \lan 0|T\{j^A j^B\}|0\ran.
\end{equation}
Moreover, the process of scattering of an electron by an electron,
\begin{equation}
    C=C_{\bar{\al}\bar{\be}\be\al}a^\dag_{\bar{\al}} a^\dag_{\bar{\be}}  a_{\be}a_{\al},\qquad C_{\bar{\al}\bar{\be}\be\al}=-\frac12 \lan\bar{\be},e^-;\bar{\al},e^-|:j^A j^B:|\al,e^-;\be,e^-\ran \lan0|T\{A_A A_B\}|0\ran,
\end{equation}
is also of the second order.

Then the expansion of the $S$-matrix with respect to the coupling constant is given by
\begin{equation}\label{S_matrix}
    S=1+V+E+B+\Pi+M+C+\cdots.
\end{equation}
The inclusive probability to record a photon in the states specified by the projector $D$ becomes
\begin{equation}\label{inclus_prob_PD}
    P_D=\Sp(PSRS^\dag).
\end{equation}
where the operator $P$ is defined in \eqref{P_D1}. As it has been already mentioned, the intensity of radiation, $I_D$, is a linear part of \eqref{inclus_prob_PD} with respect to $D$.

\subsubsection{Contributions to the inclusive probability}

The leading term of the perturbation series reads
\begin{equation}\label{leading_term}
    \Sp(PR)=\Sp(P_{ph} R_{ph}).
\end{equation}
For the coherent state of photons \eqref{coher_state}, it takes the form
\begin{equation}\label{background_coh}
    \Sp(P_{ph} R_{ph})=1-e^{-\bar{d}Dd}\approx\bar{d}Dd.
\end{equation}
As for the thermal state of photons, it turns out to be
\begin{equation}\label{background_therm}
    \Sp(P_{ph} R_{ph})=1-\frac{1}{\det(1+n_b D)}\approx\Sp(n_bD).
\end{equation}
Hereinafter, the approximate equalities for traces mean that only linear in $D$ terms are retained.

At the first order in the coupling constant, the inclusive probability \eqref{inclus_prob_PD} contains the terms
\begin{equation}
    \Sp(PVR)+c.c.
\end{equation}
The first term can be written as
\begin{equation}\label{SpPVR}
    \Sp(PVR)=N\rho^{(N,1)}_{\al\bar{\al}}\big[V^{\bar{\ga}}_{\bar{\al}\al}\Sp(P_{ph} c^\dag_{\bar{\ga}} R_{ph}) -V^{\dag\ga}_{\bar{\al}\al}\Sp(P_{ph} c_\ga R_{ph})\big],
\end{equation}
where $\rho^{(N,1)}_{\al\bar{\al}}$ is the one-particle density matrix.

Recall that if the trace-class operator $F$ is irreducible under the adjoint action of a unitary symmetry operator $\exp(i\la Q)$, i.e.,
\begin{equation}\label{irred_cond_abel}
    e^{i\la Q}Fe^{-i\la Q}=e^{i q_F\la}F,\qquad\forall\la\in\R,
\end{equation}
and $q_F\neq0$, then
\begin{equation}\label{zero_trace}
    \Sp F=0.
\end{equation}
In the differential form, the condition \eqref{irred_cond_abel} becomes
\begin{equation}
    [Q,F]=q_FF.
\end{equation}
Using this property and bearing in mind that $P_{ph}$ commutes with the photon particle number operator, we see that the contribution \eqref{SpPVR} vanishes for the initial state $R_{ph}$ commuting with the photon particle number operator.

The most important example of the initial state that does not commute with the particle number operator is the coherent state of photons \eqref{coher_state}. In this case,
\begin{equation}
\begin{split}
    \Sp(P_{ph} c^\dag_{\bar{\ga}} R_{ph})&=e^{-\bar{d}d}\lan\bar{d}|P_{ph} c^\dag_\ga|d\ran=e^{-\bar{d}d}\frac{\de}{\de d_{\bar{\ga}}} \big[e^{\bar{d}d} (1-e^{-\bar{d}Dd})\big]=\\
    &=(\bar{d}D)_{\bar{\ga}} +(1-e^{-\bar{d}Dd})(\bar{d}\tilde{D})_{\bar{\ga}}\approx (\bar{d}D)_{\bar{\ga}} +(\bar{d}Dd) \bar{d}_{\bar{\ga}},\\
    \Sp(P_{ph} c_\ga R_{ph})&=(1-e^{-\bar{d}Dd}) d_\ga\approx(\bar{d}Dd)d_\ga.
\end{split}
\end{equation}
As a result, taking into account only the first two contributions to \eqref{S_matrix}, we have
\begin{equation}\label{leading_ord}
\begin{split}
    P_D&=(1-e^{-\bar{d}Dd})\big[1-N\rho^{(N,1)}_{\al\bar{\al}} (V^{\bar{\ga}}_{\bar{\al}\al} (\bar{d}D)_{\bar{\ga}} + V^{\dag \ga}_{\bar{\al}\al} (Dd)_{\ga})\big] +N\rho^{(N,1)}_{\al\bar{\al}} (V^{\bar{\ga}}_{\bar{\al}\al} (\bar{d}D)_{\bar{\ga}} + V^{\dag \ga}_{\bar{\al}\al} (Dd)_{\ga}),\\
    I_D&= (\bar{d}Dd)+N\rho^{(N,1)}_{\al\bar{\al}} (V^{\bar{\ga}}_{\bar{\al}\al} (\bar{d}D)_{\bar{\ga}} + V^{\dag \ga}_{\bar{\al}\al} (Dd)_{\ga}).
\end{split}
\end{equation}
The contribution of unity in the square brackets on the first line describes the inclusive probability to record a photon in the initial beam of photons, whereas the whole expression in the square brackets describes a change (renormalization) of the background radiation. For the sufficiently small amplitude $|d_\ga|$, the expressions for $I_D$ and $P_D$ are formally the same. The explicit expressions for $\rho^{(N,1)}_{\al\bar{\al}}$ and $V^{\bar{\ga}}_{\bar{\al}\al}$ have been already given above.

Notice that, in the coherent state, the average electromagnetic potential reads
\begin{equation}
    A^c_A=e_A^\ga d_\ga+ \bar{e}_A^{\bar{\ga}} \bar{d}_{\bar{\ga}},
\end{equation}
and the average electric field strength has the form
\begin{equation}
    E^c_A=-i(k_{0\ga}e_A^\ga d_\ga- k_{0\bar{\ga}}\bar{e}_A^{\bar{\ga}} \bar{d}_{\bar{\ga}}).
\end{equation}
In fact, these expressions are not zero only for the spatial values of the index $\mu$ entering into $A$.

Let us define
\begin{equation}
    E^c_{\ga A}:=-i(k_{0\ga}e_A^\ga d_\ga- k_{0 \ga}\bar{e}_A^{\ga} \bar{d}_{\ga}),
\end{equation}
where there is no summation over $\ga$ on the right-hand side. Then taking $D_{\al\bar{\al}}=k_{0\ga}\pr^\ga_{\al\bar{\al}}$, we obtain
\begin{equation}
    V^{\bar{\de}}_{\bar{\al}\al} (\bar{d}D)_{\bar{\de}} + V^{\dag \de}_{\bar{\al}\al} (Dd)_{\de}=-\lan\bar{\al},e^-|j^A|\al,e^-\ran E^c_{\ga A}.
\end{equation}
In particular, for the pure initial state of a single electron, the intensity of radiation is determined by the diagonal of the current operator, i.e., by the Dirac current of the particle wave packet \cite{PanGov18,GovPan18,PanGov19,PanGov21},
\begin{equation}
    N\rho^{(N,1)}_{\al\bar{\al}} (V^{\bar{\de}}_{\bar{\al}\al} (\bar{d}D)_{\bar{\de}} + V^{\dag \de}_{\bar{\al}\al} (Dd)_{\de})=k_{0\ga}(\mathcal{A}^\ga_{cl}\bar{d}_\ga+c.c.)\qquad\text{(no summation over $\ga$)}.
\end{equation}
The quantity $\mathcal{A}^{\bar{\ga}}_{cl}$ is the amplitude of radiation of a photon with the quantum numbers $\bar{\ga}$ by the classical Dirac current. In the leading (first) order in the coupling constant, it is given by \cite{Glaub2,KlauSud,ippccb}
\begin{equation}\label{class_amplit}
    \mathcal{A}^{\bar{\ga}}_{cl}:=-i\rho^{(1,1)}_{\al\bar{\al}}\lan\bar{\al},e^-|j^A|\al,e^-\ran\bar{e}_A^{\bar{\ga}}=-ie\int dx\bar{\vf}(x)\ga^\mu\vf(x)  \frac{e_\mu^{*(\la)}(\spk)}{\sqrt{2k_0 V}}e^{ik_\nu x^\nu},
\end{equation}
where, in the last equality, it is assumed just for definiteness that the mode functions of photons are the plane waves in a vacuum and
\begin{equation}
    \vf(x)=\sum_\al \vf_\al\lan0| \psi(x)|\al,e^-\ran=\sum_s\int\frac{V d\spp}{(2\pi)^3}\sqrt{\frac{m}{V p_0}} u_s(\spp)\vf_s(\spp)e^{-ip_\nu x^\nu}.
\end{equation}
In the last equality, it is supposed that the evolution of a quantum Dirac field obeys the free Dirac equation. If the quantum field satisfies the Dirac equation in the external field, then the mode functions of the Dirac field and $\vf(x)$ must be the solutions of the Dirac equation in this field.

The abovementioned property of the contribution to the inclusive probability we consider can be used for probing the form of the wave packet by means of its stimulated radiation. To this end, one can apply the developed techniques for diagnostics of beams of charged particles \cite{Pbook,SukhKubPot17,SukhihDTHs}. Furthermore, one can amplify stimulated radiation from a wave packet at certain energy harmonics by modulating the profile of this wave packet, for example, by endowing it with some periodic structure. We shall consider this contribution to the inclusive probability to record a photon in detail in Sec. \ref{Stim_Rad}.

In the next order of perturbation theory for $P_D$, there are the following contributions
\begin{equation}\label{second_ord_trms}
    \Sp(PVRV^\dag)+\Sp(PERE^\dag)+[\Sp(PBR)+\Sp(P\Pi R)+\Sp(PMR)+\Sp(PCR)+c.c.].
\end{equation}
Consider the first term. Substituting the explicit expression for $V$, it is not difficult to deduce that
\begin{equation}
\begin{split}
    \Sp(PVRV^\dag)=&\,\Sp(P_{ph} c^\dag_{\bar{\ga}} R_{ph} c_{\ga})\big[N\rho^{(N,1)}_{\al\bar{\al}}  V^{\dag\ga}_{\bar{\al}\be} V^{\bar{\ga}}_{\be\al} +N(N-1)\rho^{(N,2)}_{\al_2\al_1|\bar{\al}_1\bar{\al}_2} V^{\dag\ga}_{\bar{\al}_2\al_2} V^{\bar{\ga}}_{\bar{\al}_1\al_1} \big]+\\
    &+\Sp(P_{ph} c_{\ga} R_{ph} c^\dag_{\bar{\ga}})\big[N\rho^{(N,1)}_{\al\bar{\al}}  V^{\bar{\ga}}_{\bar{\al}\be} V^{\dag\ga}_{\be\al} +N(N-1)\rho^{(N,2)}_{\al_2\al_1|\bar{\al}_1\bar{\al}_2} V^{\bar{\ga}}_{\bar{\al}_2\al_2} V^{\dag\ga}_{\bar{\al}_1\al_1} \big]-\\
    &-\Sp(P_{ph} c^\dag_{\bar{\ga}_1} R_{ph} c^\dag_{\bar{\ga}_2})\big[N\rho^{(N,1)}_{\al\bar{\al}}  V^{\bar{\ga}_1}_{\bar{\al}\be} V^{\bar{\ga}_2}_{\be\al} +N(N-1)\rho^{(N,2)}_{\al_2\al_1|\bar{\al}_1\bar{\al}_2} V^{\bar{\ga}_1}_{\bar{\al}_2\al_2} V^{\bar{\ga}_2}_{\bar{\al}_1\al_1} \big]-\\
    &-\Sp(P_{ph} c_{\ga_1} R_{ph} c_{\ga_2})\big[N\rho^{(N,1)}_{\al\bar{\al}}  V^{\dag\ga_1}_{\bar{\al}\be} V^{\dag\ga_2}_{\be\al} +N(N-1)\rho^{(N,2)}_{\al_2\al_1|\bar{\al}_1\bar{\al}_2} V^{\dag\ga_1}_{\bar{\al}_2\al_2} V^{\dag\ga_2}_{\bar{\al}_1\al_1} \big].
\end{split}
\end{equation}
The explicit expressions for $\rho^{(N,2)}_{\al_2\al_1|\bar{\al}_1\bar{\al}_2}$ were given above. This formula is the generalization of formula (A.15) of the paper \cite{pra103} to the case of the initial state of the system of the form \eqref{dens_matr_ini}. As far as the coherent initial state of photons \eqref{coher_state} is concerned, we have
\begin{equation}\label{trace_V}
\begin{split}
    \Sp(P_{ph} c^\dag_{\bar{\ga}} R_{ph} c_{\ga})&=\de_{\ga\bar{\ga}}+ d_{\ga}\bar{d}_{\bar{\ga}} -(\tilde{D}_{\ga\bar{\ga}} +(\tilde{D}d)_{\ga} (\bar{d}\tilde{D})_{\bar{\ga}} )e^{-\bar{d}Dd}\approx\\
    &\approx (\de_{\ga\bar{\ga}}+d_{\ga}\bar{d}_{\bar{\ga}}
    )(\bar{d}Dd) +D_{\ga\bar{\ga}} +(Dd)_{\ga} \bar{d}_{\bar{\ga}} +d_{\ga} (\bar{d}D)_{\bar{\ga}},\\
    \Sp(P_{ph} c_{\ga}R_{ph} c^\dag_{\bar{\ga}})&=d_{\ga} \bar{d}_{\bar{\ga}}(1-e^{-\bar{d}Dd})\approx d_{\ga} \bar{d}_{\bar{\ga}} (\bar{d}Dd),\\
    \Sp(P_{ph} c^\dag_{\bar{\ga}_1}R_{ph} c^\dag_{\bar{\ga}_2})&=\bar{d}_{\bar{\ga}_1} \bar{d}_{\bar{\ga}_2}(1-e^{-\bar{d}Dd}) +(\bar{d}D)_{\bar{\ga}_1}\bar{d}_{\bar{\ga}_2}e^{-\bar{d}Dd} \approx \bar{d}_{\bar{\ga}_1} \bar{d}_{\bar{\ga}_2} (\bar{d}Dd) +(\bar{d}D)_{\bar{\ga}_1}\bar{d}_{\bar{\ga}_2},\\
    \Sp(P_{ph} c_{\ga_1}R_{ph} c_{\ga_2})&= d_{\ga_1} d_{\ga_2} (1-e^{-\bar{d}Dd}) +d_{\ga_1} (Dd)_{\ga_2} e^{-\bar{d}Dd}\approx d_{\ga_1} d_{\ga_2} (\bar{d}Dd) +d_{\ga_1} (Dd)_{\ga_2}.
\end{split}
\end{equation}
In the approximate expressions, the contributions proportional to $(\bar{d}Dd)$ are responsible for renormalization of the intensity of the background radiation.

Let
\begin{equation}
\begin{gathered}
    V_{D\bar{\al}\be}:=V_{\bar{\al}\be}^{\bar{\ga}} (\bar{d}D)_{\bar{\ga}},\qquad V^\dag_{D\bar{\al}\be}:=V_{\bar{\al}\be}^{\dag\ga} (Dd)_{\ga},\\
    v_{\bar{\al}\be}:=V_{\bar{\al}\be}^{\bar{\ga}} \bar{d}_{\bar{\ga}} -V_{\bar{\al}\be}^{\dag\ga} d_{\ga},\qquad v_{D\bar{\al}\be}:=V_{D\bar{\al}\be}-V^\dag_{D\bar{\al}\be}.
\end{gathered}
\end{equation}
Notice that
\begin{equation}
    v_{\bar{\al}\be}=-i\lan\bar{\al},e^-|j^A|\be,e^-\ran A^c_A.
\end{equation}
Then
\begin{equation}
\begin{split}
    \Sp(PVRV^\dag)=&\,N\rho^{(N,1)}_{\al\bar{\al}}  \Big\{\big[D_{\ga\bar{\ga}}+\tilde{D}_{\ga\bar{\ga}}(1-e^{-\bar{d}Dd})\big]V^{\dag\ga}_{\bar{\al}\be} V^{\bar{\ga}}_{\be\al} -v_{\bar{\al}\be} v_{\be\al} (1-e^{-\bar{d}Dd})+\\
    &+(V^\dag_{D\bar{\al}\be}v_{\be\al} -v_{\bar{\al}\be} V_{D\be\al}-V^\dag_{D\bar{\al}\be}V_{D\be\al})e^{-\bar{d}Dd}\Big\}+\\ &+N(N-1)\rho^{(N,2)}_{\al_2\al_1|\bar{\al}_1\bar{\al}_2}\Big\{\big[D_{\ga\bar{\ga}} +\tilde{D}_{\ga\bar{\ga}}(1-e^{-\bar{d}Dd})\big]V^{\dag\ga}_{\bar{\al}_2\al_2} V^{\bar{\ga}}_{\bar{\al}_1\al_1}
    -v_{\bar{\al}_2\al_2} v_{\bar{\al}_1\al_1} (1-e^{-\bar{d}Dd})-\\
    &-(v_{\bar{\al}_2\al_2}v_{D\bar{\al}_1\al_1} +V^\dag_{D\bar{\al}_2\al_2}V_{D\bar{\al}_1\al_1})e^{-\bar{d}Dd} \Big\}\approx\\
    \approx&\,N\rho^{(N,1)}_{\al\bar{\al}}  \Big\{\big[D_{\ga\bar{\ga}}+\de_{\ga\bar{\ga}}(\bar{d}Dd)\big]V^{\dag\ga}_{\bar{\al}\be} V^{\bar{\ga}}_{\be\al} -v_{\bar{\al}\be} v_{\be\al} (\bar{d}Dd)+V^\dag_{D\bar{\al}\be}v_{\be\al} -v_{\bar{\al}\be} V_{D\be\al}\Big\}+\\ &+N(N-1)\rho^{(N,2)}_{\al_2\al_1|\bar{\al}_1\bar{\al}_2} \Big\{\big[D_{\ga\bar{\ga}} +\de_{\ga\bar{\ga}}(\bar{d}Dd)\big]V^{\dag\ga}_{\bar{\al}_2\al_2} V^{\bar{\ga}}_{\bar{\al}_1\al_1}
    -v_{\bar{\al}_2\al_2} v_{\bar{\al}_1\al_1} (\bar{d}Dd)-v_{\bar{\al}_2\al_2}v_{D\bar{\al}_1\al_1} \Big\}.
\end{split}
\end{equation}
In the case of small amplitudes $|d_\ga|$, in particular, for the vacuum initial state of photons, the expression is drastically simplified
\begin{equation}\label{SpPVRV}
    \Sp(PVRV^\dag)=N\rho^{(N,1)}_{\al\bar{\al}}D_{\ga\bar{\ga}}V^{\dag\ga}_{\bar{\al}\be} V^{\bar{\ga}}_{\be\al} +N(N-1)\rho^{(N,2)}_{\al_2\al_1|\bar{\al}_1\bar{\al}_2} D_{\ga\bar{\ga}}V^{\dag\ga}_{\bar{\al}_2\al_2} V^{\bar{\ga}}_{\bar{\al}_1\al_1}.
\end{equation}
This formula was obtained in \cite{pra103}, where the main properties of this expression were also discussed. In particular, the last term in \eqref{SpPVRV} describes the contribution of coherent radiation produced by the beam of particles. On neglecting the exchange term, it is determined by the classical Dirac currents created by the wave packets of particles constituting the beam as in formula \eqref{class_amplit}. The first term in \eqref{SpPVRV} describes incoherent radiation from the beam of particles. The quantum recoil due to radiation of a photon is essential for this contribution and so in many cases the forms of the wave packets of particles constituting the beam turn out to be irrelevant to for the inclusive probability to a much extent \cite{MarcuseI,MarcuseII,SundMil90,LapFedEber93,PMHK08,CorsPeat11,WCCP16,AngDiPiaz18,Remez19,KdGdAR21,Wong21,BBTq1,BBTq2}. Notice that, in the case when the initial state does not contain photons, only the contribution \eqref{SpPVRV} to the inclusive probability does not vanish for those orders of the perturbation theory that we study. In other words, this contribution is responsible for spontaneous radiation created by the beam of particles.

Now we consider the thermal initial state of photons. In virtue of the property \eqref{zero_trace}, only two of the traces \eqref{trace_V} are different from zero for the thermal state of photons because the density matrix $R_{ph}$ of such a state commutes with the particle number operator. The nonvanishing traces are
\begin{equation}\label{tr_therm}
\begin{split}
    \Sp(P_{ph} c^\dag_{\bar{\ga}} R_{ph} c_{\ga})&=(1+n_b)_{\ga\bar{\ga}}-\frac{1}{\det(1+n_bD)}[\tilde{D}(1+n_bD)^{-1}(1+n_b)]_{\ga\bar{\ga}}\approx\\
    &\approx (1+n_b)_{\ga\bar{\ga}}\Sp(n_bD)+[(1+n_b)D(1+n_b)]_{\ga\bar{\ga}},\\
    \Sp(P_{ph} c_{\ga}R_{ph} c^\dag_{\bar{\ga}})&=(n_b)_{\ga\bar{\ga}} -\frac{1}{\det(1+n_bD)}[(1+n_bD)^{-1}n_b]_{\ga\bar{\ga}}\approx\\
    &\approx (n_b)_{\ga\bar{\ga}}\Sp(n_bD)+(n_bDn_b)_{\ga\bar{\ga}}.
\end{split}
\end{equation}
The second term on the second line describes the influence of the presence of a photon gas at finite temperature to the radiation from charged particles. This contribution was obtained in formula (50) of the paper \cite{BKL5}. The first term on this line and the first term on the fourth line of \eqref{tr_therm} are responsible for renormalization of the background thermal radiation. The second term on the fourth line of \eqref{tr_therm} describes the effect of absorption of photons by the radiating system on the intensity of radiation. As a result,
\begin{align*}
%\begin{split}
    \Sp(PVRV^\dag)=&\Big\{(1+n_b)_{\ga\bar{\ga}}-\frac{1}{\det(1+n_bD)}[\tilde{D}(1+n_bD)^{-1}(1+n_b)]_{\ga\bar{\ga}} \Big\}\times\\
    &\times \Big[N\rho^{(N,1)}_{\al\bar{\al}}V^{\dag\ga}_{\bar{\al}\be} V^{\bar{\ga}}_{\be\al} +N(N-1)\rho^{(N,2)}_{\al_2\al_1|\bar{\al}_1\bar{\al}_2}V^{\dag\ga}_{\bar{\al}_2\al_2} V^{\bar{\ga}}_{\bar{\al}_1\al_1}  \Big]+\\
    &+\Big\{(n_b)_{\ga\bar{\ga}}-\frac{1}{\det(1+n_bD)}[(1+n_bD)^{-1}n_b]_{\ga\bar{\ga}} \Big\}\times\ttg\\
    &\times \Big[N\rho^{(N,1)}_{\al\bar{\al}}V^{\bar{\ga}}_{\bar{\al}\be} V^{\dag\ga}_{\be\al} +N(N-1)\rho^{(N,2)}_{\al_2\al_1|\bar{\al}_1\bar{\al}_2}V^{\bar{\ga}}_{\bar{\al}_2\al_2} V^{\dag\ga}_{\bar{\al}_1\al_1}  \Big]\approx\\
    \approx&\big\{(1+n_b)_{\ga\bar{\ga}}\Sp(n_bD)+[(1+n_b)D(1+n_b)]_{\ga\bar{\ga}}\big\}\times\\
    &\times \Big[N\rho^{(N,1)}_{\al\bar{\al}}V^{\dag\ga}_{\bar{\al}\be} V^{\bar{\ga}}_{\be\al} +N(N-1)\rho^{(N,2)}_{\al_2\al_1|\bar{\al}_1\bar{\al}_2}V^{\dag\ga}_{\bar{\al}_2\al_2} V^{\bar{\ga}}_{\bar{\al}_1\al_1}  \Big]+\\
    &+\big\{(n_b)_{\ga\bar{\ga}}\Sp(n_bD)+[n_bDn_b]_{\ga\bar{\ga}} \big\}\times\\
    &\times \Big[N\rho^{(N,1)}_{\al\bar{\al}}V^{\bar{\ga}}_{\bar{\al}\be} V^{\dag\ga}_{\be\al} +N(N-1)\rho^{(N,2)}_{\al_2\al_1|\bar{\al}_1\bar{\al}_2}V^{\bar{\ga}}_{\bar{\al}_2\al_2} V^{\dag\ga}_{\bar{\al}_1\al_1}  \Big]
%\end{split}
\end{align*}
The last term in this expression, standing on the seventh and eighth lines, which is responsible for the absorption of thermal photons by the system, was missed in \cite{BKL5}.

As for the second term in \eqref{second_ord_trms}, we obtain
\begin{equation}\label{SpPERE}
    \Sp(PERE^\dag)=\Sp(P_{ph} c_{\ga}R_{ph} c^\dag_{\bar{\ga}})(E^{\dag\bar{\ga}}_{\al\be} E^{\ga}_{\al\be}-N\rho^{(N,1)}_{\bar{\al}\al} E^{\dag\bar{\ga}}_{\bar{\al}\be} E^{\ga}_{\al\be}).
\end{equation}
The explicit expressions for the first factor in \eqref{SpPERE} were derived in \eqref{trace_V}, \eqref{tr_therm}.

The third term in \eqref{second_ord_trms} is written as
\begin{equation}\label{SpPBR0}
    \Sp(PBR)=N\rho^{(N,1)}_{\al\bar{\al}}\big[ B^{\bar{\ga}\ga}_{1\bar{\al}\al} \Sp(P_{ph} c^\dag_{\bar{\ga}}c_{\ga}R_{ph}) +B^{\ga_1\ga_2}_{2\bar{\al}\al} \Sp(P_{ph} c_{\ga_1} c_{\ga_2}R_{ph}) +B^{\bar{\ga}_1\bar{\ga}_2}_{3\bar{\al}\al} \Sp(P_{ph} c^\dag_{\bar{\ga}_1} c^\dag_{\bar{\ga}_2}R_{ph})\big].
\end{equation}
In the case of the coherent initial state of photons \eqref{coher_state}, we come to
\begin{equation}
\begin{split}
    \Sp(P_{ph} c^\dag_{\bar{\ga}}c_{\ga}R_{ph})&=(1-e^{-\bar{d}Dd})(\bar{d}\tilde{D})_{\bar{\ga}} d_{\ga}+(\bar{d}D)_{\bar{\ga}} d_{\ga} \approx \bar{d}_{\bar{\ga}} d_{\ga}(\bar{d}Dd)+(\bar{d}D)_{\bar{\ga}} d_{\ga},\\
    \Sp(P_{ph} c_{\ga_1}c_{\ga_2}R_{ph})&=(1-e^{-\bar{d}Dd})d_{\ga_1}d_{\ga_2}\approx d_{\ga_1}d_{\ga_2}(\bar{d}Dd),\\
    \Sp(P_{ph} c^\dag_{\bar{\ga}_1}c^\dag_{\bar{\ga}_2}R_{ph})&=\bar{d}_{\bar{\ga}_1}\bar{d}_{\bar{\ga}_2} -(\bar{d}\tilde{D})_{\bar{\ga}_1}(\bar{d}\tilde{D})_{\bar{\ga}_2} e^{-\bar{d}Dd}\approx \bar{d}_{\bar{\ga}_1}\bar{d}_{\bar{\ga}_2} (\bar{d}Dd) +(\bar{d}D)_{\bar{\ga}_1}\bar{d}_{\bar{\ga}_2} +\bar{d}_{\bar{\ga}_1}(\bar{d}D)_{\bar{\ga}_2}.
\end{split}
\end{equation}
Substituting the explicit expressions for $B_{1,2,3}$, we deduce
\begin{equation}\label{SpPBR}
\begin{split}
    \Sp(PBR)&=N\rho^{(N,1)}_{\al\bar{\al}}B^{AB}_{\bar{\al}\al}\Big\{(1-e^{-\bar{d}Dd}) [(e_A d)+(\bar{d}\tilde{D}\bar{e}_A)] [(e_B d)+(\bar{d}\tilde{D}\bar{e}_B)] +(\bar{d}D\bar{e}_A)[2A^c_B -(\bar{d}D\bar{e}_B)] \Big\}\approx\\
    &\approx N\rho^{(N,1)}_{\al\bar{\al}}B^{AB}_{\bar{\al}\al}\big[A^c_A A^c_B (\bar{d}Dd) +2(\bar{d}D\bar{e}_A)A^c_B \big].
\end{split}
\end{equation}
In the case of the thermal initial state of photons, the nonvanishing trace is
\begin{equation}\label{SpPccR_th}
\begin{split}
    \Sp(P_{ph} c^\dag_{\bar{\ga}}c_{\ga}R_{ph})&=(n_b)_{\ga\bar{\ga}}-\frac{1}{\det(1+n_bD)}[n_b(1+Dn_b)^{-1}\tilde{D}]_{\ga\bar{\ga}}\approx\\
    &\approx (n_b)_{\ga\bar{\ga}}\Sp(n_bD) +[n_bD(1+n_b)]_{\ga\bar{\ga}}.
\end{split}
\end{equation}
The other traces entering into \eqref{SpPBR} vanish. Thus,
\begin{equation}\label{SpPBRth}
\begin{split}
    \Sp(PBR)&=2N\rho^{(N,1)}_{\al\bar{\al}}B^{AB}_{\bar{\al}\al}\Big[(e_Bn_b\bar{e}_A) -\frac{1}{\det(1+n_bD)}(e_Bn_b(1+Dn_b)^{-1}\tilde{D}\bar{e}_A) \Big]\approx\\
    &\approx2N\rho^{(N,1)}_{\al\bar{\al}}B^{AB}_{\bar{\al}\al}\big[ (e_Bn_b\bar{e}_A)\Sp(n_bD) +(e_Bn_bD(1+n_b)\bar{e}_A)\big].
\end{split}
\end{equation}
The interpretation of the terms on the last line in the same as for the contributions considered above. The first term describes renormalization of the intensity of the background thermal radiation, while the second term describes the intensity of reradiation of thermal photons by electrons in the leading order of perturbation theory.

In the same way, the contribution of the polarization operator is evaluated. For the coherent initial state of photons, we arrive at
\begin{equation}\label{SpPPiR}
\begin{split}
    \Sp(P\Pi R)&=\Pi^{AB} \Big\{(1-e^{-\bar{d}Dd}) [(e_A d)+(\bar{d}\tilde{D}\bar{e}_A)] [(e_B d)+(\bar{d}\tilde{D}\bar{e}_B)] +(\bar{d}D\bar{e}_A)[2A^c_B -(\bar{d}D\bar{e}_B)] \Big\}\approx\\
    &\approx \Pi^{AB}\big[A^c_A A^c_B (\bar{d}Dd) +2(\bar{d}D\bar{e}_A)A^c_B \big].
\end{split}
\end{equation}
If the initial state of photons is thermal, then
\begin{equation}\label{SpPPiRth}
\begin{split}
    \Sp(P\Pi R)&=2\Pi^{AB}\Big[(e_Bn_b\bar{e}_A) -\frac{1}{\det(1+n_bD)}(e_Bn_b(1+Dn_b)^{-1}\tilde{D}\bar{e}_A) \Big]\approx\\
    &\approx2\Pi^{AB}\big[ (e_Bn_b\bar{e}_A)\Sp(n_bD) +(e_Bn_bD(1+n_b)\bar{e}_A)\big].
\end{split}
\end{equation}
By comparing \eqref{SpPBR}, \eqref{SpPBRth} with \eqref{SpPPiR}, \eqref{SpPPiRth}, we see that the quantity,
\begin{equation}\label{Pi_eff}
    \Pi^{AB}_{eff}:=\Pi^{AB}+N\rho^{(N,1)}_{\al\bar{\al}}B^{AB}_{\bar{\al}\al},
\end{equation}
plays the role of an effective susceptibility tensor. The second term describes the correction to the vacuum susceptibility tensor caused by the presence of electrons in the initial state. This correction is different from zero even in the case $N=1$. In the absence of the external field, the contribution $\Pi^{AB}$ is removed by renormalization of the electromagnetic field strength. Then $\Pi^{AB}_{eff}$ is determined by the second term in \eqref{Pi_eff}. We shall consider this contribution in more detail in Sec. \ref{Electr_Suscept}.

The fifth and sixth terms in \eqref{second_ord_trms} give
\begin{equation}\label{self_inter}
    \Sp(PMR)+\Sp(PCR)=\big[N\rho^{(N,1)}_{\al\bar{\al}} M_{\bar{\al}\al} +N(N-1)\rho^{(N,2)}_{\al_2\al_1|\bar{\al}_1\bar{\al}_2} C_{\bar{\al}_1\bar{\al}_2\al_2\al_1}\big] \Sp(P_{ph} R_{ph}).
\end{equation}
These contributions lead only to renormalization of the intensity of the background radiation. The quantity in the square brackets describes the correction to the self-energy of the beam of particles.

\subsection{Background electromagnetic field method}

Let us consider the case when the initial state of photons is coherent, \eqref{coher_state}, and the electromagnetic field corresponding to this state is taken into account nonperturbatively. It is known \cite{Ritus,PiMuHaKermp,IldertRev22} that the external electromagnetic field can be treated perturbatively provided the so-called undulator strength parameter
\begin{equation}
    |e a_0/m|\ll1,
\end{equation}
where $a_0$ is the typical value of the electromagnetic potential in the Coulomb gauge in the laboratory frame. If the above condition is not fulfilled, one needs to employ the background field method accounting for the external field nonperturbatively.

Let us introduce the unitary displacement operator
\begin{equation}
    \mathcal{D}(d):=e^{c^\dag d-\bar{d}c}.
\end{equation}
Then the initial coherent state of photons is
\begin{equation}
    \mathcal{D}(d)|0\ran_{ph}.
\end{equation}
The average of the operator $O$ in the Schr\"{o}dinger representation is given by
\begin{equation}
    \lan O\ran=\Sp[OU\mathcal{D}(d)\big(|0\ran_{ph}\lan0|_{ph}\otimes R_e\otimes|0\ran_{e^+}\lan0|_{e^+} \big)\mathcal{D}^\dag(d)U^\dag],
\end{equation}
where $U$ is the evolution operator over an infinite time. This expression can be rewritten as
\begin{equation}
    \lan O\ran=\Sp(O_dU_d R U^\dag_d),
\end{equation}
where
\begin{equation}
    R:=|0\ran_{ph}\lan0|_{ph}\otimes R_e\otimes|0\ran_{e^+}\lan0|_{e^+} ,\qquad U_d:=\mathcal{D}^\dag(d)U\mathcal{D}(d),\qquad O_d:= \mathcal{D}^\dag(d)O\mathcal{D}(d).
\end{equation}
It is clear that
\begin{equation}
    \mathcal{D}^\dag(d) \hat{A}_A \mathcal{D}(d)= \hat{A}_A +A^c_A.
\end{equation}
Moreover,
\begin{equation}
    P^d:=\mathcal{D}^\dag(d) P \mathcal{D}(d)=P^d_{ph}\otimes1_{e} \otimes1_{e^+},
\end{equation}
where
\begin{equation}
    P^d_{ph}=1-e^{-\bar{d}Dd}e^{-c^\dag Dd}:e^{-c^\dag Dc}:e^{-\bar{d}Dc}.
\end{equation}
As for the action functional of quantum electrodynamics, we have
\begin{equation}\label{action_shifted}
    S[A,\bar{\psi},\psi]\rightarrow S[A+A^c,\bar{\psi},\psi]=\int d^4x \big\{\bar{\psi}(\hat{p}-e\hat{A}^c-m)\psi -\frac14 f_{\mu\nu}f^{\mu\nu} -e\bar{\psi}\hat{A}\psi -\frac{1}{4}F_{\mu\nu}F^{\mu\nu} -\frac12F_{\mu\nu}f^{\mu\nu}\big\},
\end{equation}
where $f_{\mu\nu}=\partial_{[\mu}A_{\nu]}$ and $F_{\mu\nu}=\partial_{[\mu}A^c_{\nu]}$. As long as $F_{\mu\nu}$ satisfies the free Maxwell equations,
\begin{equation}
    \partial^\nu F_{\mu\nu}=0,
\end{equation}
the last term in \eqref{action_shifted} is the total derivative. Therefore, in evaluating the evolution operator over an infinite time, or the $S$-matrix, under the standard assumption that the interaction is adiabatically switched off for $|t|\rightarrow\infty$, one can disregard this term. The penultimate term in \eqref{action_shifted} is a real number and it does not affect the averages. As a result, we arrive at the action functional of quantum electrodynamics in the Furry picture with the external field $A^c_A$. In order to take into account this field, one has to construct a complete set of solutions to the Dirac equation in such an external field (see for details \cite{GrMuRaf,GFSh.3,ippccb}).

Thus,
\begin{equation}
    \Sp(P^d_{ph} R_{ph})=1-e^{-\bar{d}Dd}\approx (\bar{d}Dd),
\end{equation}
where $R_{ph}=|0\ran_{ph}\lan0|_{ph}$. Furthermore,
\begin{equation}
    \Sp(P^dVR)=N\rho^{(N,1)}_{\al\bar{\al}} V_{D\bar{\al}\al}e^{-\bar{d}Dd}\approx N\rho^{(N,1)}_{\al\bar{\al}} V_{D\bar{\al}\al},
\end{equation}
and
\begin{equation}
    \Sp(P^dVR)+c.c.=N\rho^{(N,1)}_{\al\bar{\al}} (V_{D\bar{\al}\al}+V^\dag_{D\bar{\al}\al})e^{-\bar{d}Dd}\approx N\rho^{(N,1)}_{\al\bar{\al}}(V_{D\bar{\al}\al}+V^\dag_{D\bar{\al}\al}).
\end{equation}
Hereinafter, for brevity, the index $d$ of the operators entering into the expansion \eqref{S_matrix} is omitted.

In the next order of the perturbation theory, we obtain
\begin{equation}
\begin{split}
    \Sp(P^dVRV^\dag)=&\,\big[\de_{\ga\bar{\ga}} (1-e^{-\bar{d}Dd}) +(D_{\ga\bar{\ga}}-(Dd)_\ga (\bar{d}D)_{\bar{\ga}})e^{-\bar{d}Dd} \big]\times\\
    &\times\big[N\rho^{(N,1)}_{\al\bar{\al}}  V^{\dag\ga}_{\bar{\al}\be} V^{\bar{\ga}}_{\be\al} +N(N-1)\rho^{(N,2)}_{\al_2\al_1|\bar{\al}_1\bar{\al}_2} V^{\dag\ga}_{\bar{\al}_2\al_2} V^{\bar{\ga}}_{\bar{\al}_1\al_1} \big]\approx\\
    \approx&\,\big[\de_{\ga\bar{\ga}} (\bar{d}Dd) +D_{\ga\bar{\ga}} \big]\big[N\rho^{(N,1)}_{\al\bar{\al}}  V^{\dag\ga}_{\bar{\al}\be} V^{\bar{\ga}}_{\be\al} +N(N-1)\rho^{(N,2)}_{\al_2\al_1|\bar{\al}_1\bar{\al}_2} V^{\dag\ga}_{\bar{\al}_2\al_2} V^{\bar{\ga}}_{\bar{\al}_1\al_1} \big].
\end{split}
\end{equation}
Moreover,
\begin{equation}
    \Sp(P^dERE^\dag)=0,
\end{equation}
and
\begin{equation}
\begin{split}
    \Sp(P^dBR)=&\,-N\rho^{(N,1)}_{\al\bar{\al}}B^{\bar{\ga}_1\bar{\ga}_2}_{3\bar{\al}\al} (\bar{d}D)_{\bar{\ga}_1}(\bar{d}D)_{\bar{\ga}_2}\approx0,\\
    \Sp(P^d\Pi R)=&\,-\Pi^{\bar{\ga}_1\bar{\ga}_2}_{3} (\bar{d}D)_{\bar{\ga}_1}(\bar{d}D)_{\bar{\ga}_2}\approx0,\\
    \Sp(P^dMR)+\Sp(P^dCR)=&\,\big[N\rho^{(N,1)}_{\al\bar{\al}} M_{\bar{\al}\al} +N(N-1)\rho^{(N,2)}_{\al_2\al_1|\bar{\al}_1\bar{\al}_2} C_{\bar{\al}_1\bar{\al}_2\al_2\al_1}\big] \Sp(P^d_{ph} R_{ph}).
\end{split}
\end{equation}
The contribution on the last line boils down to renormalization of the intensity of the background radiation. As we see, the nonperturbative approach to take into account the electromagnetic field in the initial coherent state of photons simplifies considerably the contributions to the inclusive probability and to the intensity of radiation. However, it implies the knowledge of the complete set of solutions to the Dirac equation in the given electromagnetic field.

\section{Stimulated radiation from a wave packet}\label{Stim_Rad}

As the example of stimulated radiation produced by a wave packet, we consider transition radiation from the Dirac fermion traversing a mirror in the field of a laser wave. Spontaneous radiation created by the wave packet of a Dirac fermion falling onto the conducting plate was thoroughly investigated in \cite{pra103} (see also \cite{GaribYang,BazylZhev,Ginzburg,Pbook,ZarLomNer80,IvanLomon81}). In that paper, it was assumed that the mirror is ideally conducting and it is placed at $z\leqslant0$. The wave packet of the particle moves from right to left. Using formula (21) of the paper \cite{pra103}, we find
\begin{equation}
\begin{split}
    V^{\bar{\ga}}_{\bar{\al}\al}&=-i\frac{m}{V}\int dx a^*_{i\la}(x_3)\bar{u}_{\bar{\al}}\big[e\ga^i -i\mu_a(k_0\s^{i0}+(p_j-p_j')\s^{ij}) \big]u_\al\frac{e^{ik_0x^0-i\spk_\perp\spx_\perp +i(p'_\mu-p_\mu)x^\mu}}{\sqrt{2Vk_0p_0p'_0}}=\\
    &=-i\frac{m}{V}\int dx a^*_{i\la}(x_3)\bar{u}_{\bar{\al}}\big[e\ga^i -i\mu_a(p'_\nu-p_\nu)\s^{\nu i} \big]u_\al\frac{e^{ik_0x^0-i\spk_\perp\spx_\perp +i(p'_\mu-p_\mu)x^\mu}}{\sqrt{2Vk_0p_0p'_0}},
\end{split}
\end{equation}
where $\al=(s,\spp)$ and $\bar{\al}=(s',\spp')$, $\mu_a$ is the anomalous magnetic moment of the particle, and the vector $a^*_{i\la}(x_3)$ comes from the mode function of a photon with helicity $\la$ (see the notation in \cite{pra103}).

Introduce the one-particle density matrix
\begin{equation}\label{dens_matr_redef}
    \rho_{ss'}(\spp,\spp'):=\frac{V}{(2\pi)^3}\rho^{(N,1)}_{\al\bar{\al}},\qquad \sum_s\int d\spp\rho_{ss}(\spp,\spp)=1.
\end{equation}
So long as the particle impinges on the mirror from right to left, $\rho_{ss'}(\spp,\spp')$ is strongly suppressed for $p^3>0$ or $p'^{3}>0$. Employing formula (26) of the paper \cite{pra103} (there is a misprint in this formula: $\mathbf{f}_r^{(\la)}(\spk)$ should be replaced by $\mathbf{f}_r^{*(\la)}(\spk)$), the contribution to the intensity of radiation can be cast into the form
\begin{equation}\label{NrVdD}
\begin{split}
    dF_\la(\spk):=N\rho^{(N,1)}_{\al\bar{\al}}V^{\bar{\ga}}_{\bar{\al}\al}(\bar{d}D)_{\bar{\ga}}\frac{Vd\spk}{(2\pi)^3}=&-\sum_{s,s'}\int d\spp d\spp'\de(k_0+p_0'-p_0) \de(\spk_\perp+\spp'_\perp-\spp_\perp)  N\rho_{ss'}(\spp,\spp') \times\\
    &\times\frac{m}{\sqrt{p_0p_0'}}\sum_r\frac{\bar{u}^{s'}(\spp') \big[e\ga^i -i\mu_a(p'_\nu-p_\nu)\s^{\nu i} \big]u^s(\spp)}{p_3'-p_3+rk_3-i0}\frac{f_{ri}^{*(\la)}(\spk)d^*_\la(\spk)}{\sqrt{2(2\pi)^3k_0}}k_0 d\spk,
\end{split}
\end{equation}
where the normalization of the complex amplitude,
\begin{equation}\label{d_norm}
    d_\la(\spk):=d_\ga\sqrt{\frac{V}{(2\pi)^3}},
\end{equation}
is such that the intensity of the backlighting laser wave equals
\begin{equation}\label{intens_background}
    dI^0_\la(\spk)=d^*_\la(\spk) d_\la(\spk)k_0d\spk.
\end{equation}
There is no summation over $\la$ in \eqref{NrVdD}, \eqref{intens_background}. As a result, the total intensity of radiation \eqref{leading_ord} is given by
\begin{equation}\label{dI_la_stim}
    dI_\la(\spk)=dI^0_\la(\spk) +dF_\la(\spk)+dF^*_\la(\spk).
\end{equation}
Comparing the leading term in \eqref{dI_la_stim} with the correction $dF_\la(\spk)$ and also comparing this correction with the subsequent terms of the perturbation series, it is not difficult to see that the higher orders of the perturbation theory can be neglected provided the strength of the electromagnetic field created by the beam of particles is much less than the field strength of the laser wave.

In the case of a pure one-particle initial electron state, expression \eqref{NrVdD} can be written as
\begin{equation}%\label{dF_la}
    dF_\la(\spk)=\mathcal{A}_{cl}^\la(\spk)d^*_\la(\spk) k_0d\spk\qquad\text{(no summation over $\la$)},
\end{equation}
where
\begin{equation}
    \mathcal{A}_{cl}^\la(\spk)=-i\int dx e^*_i(\la,\spk;x) j^i(x),\qquad e^*_i(\la,\spk;x)= \frac{a^*_{i\la}(x_3)}{\sqrt{2(2\pi)^3k_0}}e^{ik_0x^0-i\spk_\perp\spx_\perp},
\end{equation}
and
\begin{equation}
    j^i(x)=e\bar{\vf}(x)\ga^i\vf(x)-\mu_a\partial_\nu(\bar{\vf}(x)\s^{\nu i}\vf(x)).
\end{equation}
The quantity $\mathcal{A}_{cl}^\la(\spk)$ is the amplitude of radiation of a photon with quantum numbers $(\la,\spk)$ by the classical current $j^i(x)$. The general expression for this amplitude is presented in the first equality in \eqref{class_amplit}. Notice that the contribution to the intensity of radiation we consider is of the first order in the coupling constant $e$. Therefore, up to the terms linear in $e$, the radiation intensity can be written in the form
\begin{equation}
    dI_\la(\spk)=\big|d_\la(\spk)+\mathcal{A}^\la_{cl}(\spk)\big|^2k_0d\spk.
\end{equation}
The complex amplitude $d_\la(\spk)$ can be expressed through the electromagnetic field of a laser wave as (see, e.g., \cite{ippccb})
\begin{equation}
    d_\la(\spk)=\int d\spx e^*_i(\la,\spk;x)\big[k_0A^c_i(x)+iE^c_i(x) \big],
\end{equation}
where the Coulomb gauge is implied.

As in the case of radiation created by beams of charged particles, the symmetries of the current density $j^i(x)$ result in the selection rules in the radiation we consider. Moreover, the generation of harmonics of coherent radiation is possible despite the fact that the radiation is produced by a single particle. In other words, in describing this type of radiation, one can regard the wave packet of a particle as some kind of a charged fluid.

Let us consider the most common symmetries of the one-particle density matrix. Recall that the density matrix \eqref{dens_matr_redef} is defined at the instant of time $t=0$.
\begin{enumerate}[wide, labelwidth=!, labelindent=0pt]
  \item Translations $T_a$ along the $z$ axis by an arbitrary $a$. In this case,
\begin{equation}\label{trans_symm}
    T_a\rho T^\dag_a=\rho\;\;\Leftrightarrow\;\;e^{i(p_3'-p_3)a}\rho_{ss'}(\spp,\spp')=\rho_{ss'}(\spp,\spp').
\end{equation}
Consequently,
\begin{equation}
    \rho_{ss'}(\spp,\spp')=f_{ss'}(\spp,\spp')\de(p_3-p_3').
\end{equation}
Taking into account the delta functions appearing in expression \eqref{NrVdD}, it is easy to see that the contribution to transition radiation we consider is absent in this case. In particular, this type of radiation is absent for an electron being in an ideal twisted state. Recall that such a state is an eigenstate for the translation operator along the $z$ axis \cite{BliokhVErev,LBThY}. Nevertheless, the radiation of this type is present even for a twisted electron provided that the translation symmetry along the $z$ axis is violated. This happens, for example, for wave packets of twisted electrons with Gaussian envelope.
  \item Translations $T_a$ along the $z$ axis by a fixed vector. In this case, it follows from \eqref{trans_symm} that
\begin{equation}\label{dens_matr_period}
    \rho_{ss'}(\spp,\spp')=\sum_{n=-\infty}^\infty f^n_{ss'}(\spp,\spp')\de(p_3-p_3'-qn),
\end{equation}
where $q:=2\pi/a$. Assuming that the recoil due to photon radiation is small,
\begin{equation}\label{small_recoil}
    k_0/(p_3\be_3)\ll1,
\end{equation}
we can employ formula (B1) of \cite{pra103}:
\begin{equation}
    p_3-p_3'=k_0(1-\bs{\beta}_\perp \mathbf{n}_\perp)/\be_3,
\end{equation}
where $\bs\be=\spp/p_0$. Then
\begin{equation}
    \rho_{ss'}(\spp,\spp')=\sum_{n=-\infty}^\infty f^n_{ss'}(\spp,\spp')\de\big(k_0(1-\bs{\beta}_\perp \mathbf{n}_\perp)/\be_3-qn\big).
\end{equation}
Only the terms with $n\geqslant1$ for $q>0$ and with $n\leqslant-1$ for $q<0$ are nonvanishing in this sum. Substituting this expansion into \eqref{NrVdD}, we see that, in the case of a sufficiently small dispersion of the velocity $\bs\be$ in the wave packet, the contribution to inclusive probability we consider possesses the energy harmonics that are the same as those predicted by the classical theory of coherent radiation produced by beams of charged particles (see, e.g., \cite{Pbook}). The violation of periodicity of the wave packet along the $z$ axis due to, say, Gaussian envelope gives rise to smearing of the radiation harmonics. The mention should be made that, in spite of the fact that we refer to the condition \eqref{small_recoil} as the small recoil approximation, in fact, there is no any recoil in the process we consider since this process is determined by the diagonal of the transition current. The condition \eqref{small_recoil} should be regarded as the restriction on the domain of photon energies measured in the experiment where the corresponding approximation holds.
  \item Rotations $R_\vf$ around the $z$ axis by an arbitrary angle:
\begin{equation}
    R_\vf\rho R^\dag_\vf=\rho.
\end{equation}
Substituting the left-hand side of this equality into \eqref{NrVdD}, performing the corresponding change of the integration variables $\spp$, $\spp'$, and using the covariance of the integrand of \eqref{NrVdD}, we arrive at
\begin{equation}
    \mathcal{A}_{cl}^\la(\spk)=\mathcal{A}_{cl}^\la(\spk_\vf),\quad\forall\vf\in\R,
\end{equation}
where $\spk_\vf$ is the vector $\spk$ rotated around the $z$ axis by an angle of $\vf$. Therefore, the amplitude $\mathcal{A}_{cl}^\la(\spk)$ describes the radiation of twisted photons with the projection of the total angular momentum $m=0$ \cite{BKL3,BKLb}. As a result, the intensity of radiation has the form of an interference pattern of the incident laser wave with the radiated photons with $m=0$. In the paraxial regime, $m=l+\la$, where $l$ is the projection of the orbital angular momentum, so we have $l=-\la$. If the incident laser wave is a plane wave with helicity $\la$, then the interference pattern is a twisted spiral with one arm \cite{SerboNew} since $\la=\pm1$. The chirality of this spiral is specified by the sign of $l$, i.e., in our case, by the sign of $\la$.
  \item Rotations $R_{\vf_r}$ around the $z$ axis by a fixed angle of $\vf_r=2\pi/r$, $r\in \mathbb{Z}$. In this case,
\begin{equation}
    \mathcal{A}_{cl}^\la(\spk)=\mathcal{A}_{cl}^\la(\spk_{\vf_n}).
\end{equation}
Consequently, the amplitude $\mathcal{A}_{cl}^\la(\spk)$ describes the radiation of twisted photons with the projection of the total angular momentum $m'=rk$, $k\in \mathbb{Z}$ \cite{BKL3,BKLb}.
  \item Helical symmetry:
\begin{equation}
    V=T_{\vf/q}R_\vf.
\end{equation}
For $\vf=2\pi$, this symmetry is reduced to translations considered in item 2. Hence, the one-particle density matrix takes the form \eqref{dens_matr_period} and, consequently, the radiation amplitude can be written as the sum over harmonics
\begin{equation}
    \mathcal{A}_{cl}^\la(\spk)=\sum_{n=-\infty}^\infty \mathcal{A}_n^\la(\spk).
\end{equation}
Let us assume for definiteness that the Dirac spinors appearing in \eqref{NrVdD} are the eigenvectors for the operator of the spin projection onto the $z$ axis. Then the density matrix \eqref{dens_matr_period} possesses the helical symmetry provided
\begin{equation}\label{f_n_hel}
    f^n_{ss'}(\spp_\vf,\spp'_\vf)=e^{in\vf} e^{i(s-s')\vf}f^n_{ss'}(\spp,\spp'),
\end{equation}
where $s,s'=\pm1/2$. Substituting this expression into $\mathcal{A}_n^\la(\spk_\vf)$, performing a change of integration variables $d\spp\rightarrow d\spp_\vf$, $d\spp'\rightarrow d\spp'_\vf$, and taking into account the transformation laws of spinors and gamma matrices entering into the integrand, we obtain
\begin{equation}
    \mathcal{A}_n^\la(\spk_\vf)=e^{in\vf}\mathcal{A}_n^\la(\spk).
\end{equation}
The amplitude of radiation of twisted photon with the projection of the total angular momentum $m'$ onto the $z$ axis is proportional to (see, e.g., formula (6) of \cite{parax})
\begin{equation}
    \int_{-\pi}^\pi\frac{d\vf}{2\pi}e^{-im'\vf}\mathcal{A}_{cl}^\la(\spk_\vf)=\sum_{n=-\infty}^\infty\de_{m',n}\mathcal{A}_n^\la(\spk).
\end{equation}
As a result, we deduce the selection rules,
\begin{equation}\label{hel_sel_rel}
    p_3-p_3'\approx k_0(1-\bs{\beta}_\perp \mathbf{n}_\perp)/\be_3 =qn,\qquad m'=n,
\end{equation}
coinciding with the selection rules for transition radiation from helical beams of charged particles \cite{BKLb,HemStuXiZh14}.
  \item Combination of the symmetries $4$ and $5$. In this case, the selection rules \eqref{hel_sel_rel} hold but only those harmonics $n$ are realized that are a multiple of $r$.
\end{enumerate}

In conclusion of this section, we give the explicit expression for $dF_\la(\spk)$ neglecting the recoil due to photon radiation and setting $\mu_a=0$. Using formulas (B.1), (B.4) of the paper \cite{pra103} and
\begin{equation}\label{Gord_iden}
    \bar{u}^{s'}(\spp')\ga^\mu u^s(\spp)\approx \de_{s's} p^\mu/m,
\end{equation}
we derive
\begin{equation}
    dF_\la(\spk)=\sum_s\int d\spp N\rho_{ss}(\spp,\spp') \frac{2e|\be_3|\be^i\big[k_3f^{*(\la)}_i+(q_3-k_3)\de^3_if^{*(\la)}_3\big]}{k_0\big[(1-\bs{\be}_\perp \mathbf{n}_\perp)^2-\be_3^2 n_3^2\big]} \frac{d^*_\la(\spk)}{\sqrt{2(2\pi)^3k_0}}d\spk,
\end{equation}
where
\begin{equation}
    q_\mu:=p_\mu-p'_\mu,\qquad \spp'_\perp=\spp_\perp-\spk_\perp,\qquad p_3'=p_3-k_0(1-\bs{\beta}_\perp \mathbf{n}_\perp)/\be_3.
\end{equation}
Transition and diffraction radiations from conducting (perforated) plates are used for noninvasive diagnostics of beams of charged particles \cite{Pbook,SukhKubPot17,SukhihDTHs}. As we see, the same methods can be employed for mapping the profiles of wave packets of particles.

\section{Susceptibility of an electron wave packet}\label{Electr_Suscept}

Another one process where coherent radiation of photons produced by a wave packet of a single particle is possible is the Compton scattering. Suppose that the background electromagnetic field is absent and the initial state is of the form \eqref{dens_matr_ini}, the photons being in the coherent state \eqref{coher_state} with sufficiently small amplitude $|d_\ga|$ for the perturbation theory to be applicable. In this case, the nonvanishing contributions to the inclusive probability to record a photon and to the intensity of radiation are the leading term of the perturbation series \eqref{leading_term}, the contribution of the Compton scattering \eqref{SpPBR0}, and the terms responsible for the self-interaction of the particle beam \eqref{self_inter}. In virtue of the energy-momentum conservation law, only the term with $B^{\bar{\ga}\ga}_{1\bar{\al}\al}$ is different from zero in the contribution \eqref{SpPBR0}, whereas the nonvanishing term in expression \eqref{self_inter} is the Coulomb term containing  $C_{\bar{\al}_1\bar{\al}_2\al_2\al_1}$. For $N=1$, the latter term is zero. Notice that, contrary to the process studied in the previous section, noncommutativity of the initial photon density matrix with the photon number operator is not necessary for reradiation to occur.

Let us define
\begin{equation}
    \mathcal{N}:=N\rho^{(N,1)}_{\al\bar{\al}} B_{1\bar{\al}\al}^{\bar{\de}\de}\bar{d}_{\bar{\de}}d_\de +N(N-1)\rho^{(N,2)}_{\al_2\al_1|\bar{\al}_1\bar{\al}_2} C_{\bar{\al}_1\bar{\al}_2\al_2\al_1}.
\end{equation}
Then up to the terms of the second order in the coupling constant, the intensity of radiation of photons with quantum numbers $\ga$ is written as
\begin{equation}\label{Compt_intens}
    I_\ga=k_{0\ga}\big|d_\ga\sqrt{1+\mathcal{N}}+N\rho^{(N,1)}_{\al\bar{\al}}B^{\ga\de}_{1\bar{\al}\al}d_\de \big|^2\qquad\text{(no summation over $\ga$)}.
\end{equation}
For small $|d_\ga|$ and $N=1$, one can put $\mathcal{N}=0$. The terms of the order $e^4$ in this formula can be omitted to the accuracy we work.

It was shown above that, in the case we consider, the effective susceptibility tensor is equal to
\begin{equation}\label{Pi_eff_vac}
    \Pi^{AB}_{eff}=N\rho^{(N,1)}_{\al\bar{\al}}B^{AB}_{\bar{\al}\al}=N\rho^{(N,1)}_{\al\bar{\al}}B^{AB}_{1\bar{\al}\al},
\end{equation}
where
\begin{equation}
    B^{\bar{\ga}\ga}_{1\bar{\al}\al}=2B^{AB}_{\bar{\al}\al}\bar{e}_A^{\bar{\ga}} e_B^\ga.
\end{equation}
The quantity $B^{\bar{\ga}\ga}_{1\bar{\al}\al}$ is the amplitude of the Compton scattering.

Bearing in mind our agreement for the normalization of the mode functions, we have (see, e.g., \cite{PeskSchr})
\begin{equation}\label{Compt_ampl}
    B^{\bar{\ga}\ga}_{1\bar{\al}\al}=-i(2\pi)^4\de(p+k-p'-k')\frac{e^2m e_\mu^{(\la)}(\spk)e_\nu^{*(\la')}(\spk')}{4V^2\sqrt{k_0k_0'p_0p_0'}}\bar{u}^{s'}(\spp')\Big[ \frac{\ga^\nu\hat{k}\ga^\mu +2p^\mu\ga^\nu}{pk} +\frac{\ga^\mu\hat{k}'\ga^\nu -2p^\nu\ga^\mu}{pk'} \Big]u^s(\spp).
\end{equation}
In the small quantum recoil limit, $|\De\spk|\ll E$, where $\De\spk:=\spk'-\spk$, it follows from the energy-momentum conservation law that
\begin{equation}
    k_0'-k_0=\bs{\be}(\spk'-\spk)\;\;\Leftrightarrow\;\; k_0'=k_0\frac{1-\bs{\be}\mathbf{n}}{1-\bs{\be}\mathbf{n}'}\;\;\Leftrightarrow\;\;pk'=pk.
\end{equation}
Neglecting the higher order corrections with respect to the recoil parameter, $|\De\spk|/E\ll1$, and employing the relation \eqref{Gord_iden}, the amplitude \eqref{Compt_ampl} can be cast into the form
\begin{equation}\label{B1_appr}
    B^{\bar{\ga}\ga}_{1\bar{\al}\al}=-i(2\pi)^4\de(p+k-p'-k') \frac{e^2 e_i^{(\la)}(\spk)e_i^{*(\la')}(\spk')}{2V^2 \sqrt{k_0k_0'}p_0}\de_{ss'}.
\end{equation}
Notice that, as in the previous section, there is no recoil in the process we consider. The condition $|\De\spk|/E\ll1$ should be regarded as the restriction on the domain of applicability of the approximate formula \eqref{B1_appr}. As a result, introducing the one-particle density matrix \eqref{dens_matr_redef}, we arrive at
\begin{equation}\label{NrB1}
\begin{split}
    \Phi(\la',\spk';\la,\spk)&= -2\pi i e^2 \frac{e_i^{(\la)}(\spk)e_i^{*(\la')}(\spk')}{2 V \sqrt{k_0k_0'}}\sum_s\int \frac{d\spp}{E(\spp)} \de(p_0+k_0-p'_0-k'_0)N\rho_{ss}(\spp,\spp-\De\spk)=\\
    &=-2\pi i e^2 \frac{e_i^{(\la)}(\spk)e_i^{*(\la')}(\spk')}{2 V \sqrt{k_0k_0'}}\lan e^{-i\De\spk\hat{\spx}}\de(p\De k)\ran,
\end{split}
\end{equation}
where
\begin{equation}
    \Phi(\la',\spk';\la,\spk):=N\rho^{(N,1)}_{\al\bar{\al}}B^{\bar{\ga}\ga}_{1\bar{\al}\al},
\end{equation}
and  $\hat{x}^i=i\partial/\partial p_i$.

Supposing that
\begin{equation}
    \frac{\s_e}{\s_\ga}\frac{|\De\spk|}{E}\ll1,\qquad \frac{\s_e\be}{E}\ll1,
\end{equation}
where $\s_e$ is the characteristic scale of variation of the electron wave packet in the momentum space and $\s_\ga$ is the same quantity for the incident photon wave packet, the delta function can be taken outside the average sign
\begin{equation}
    \Phi(\la',\spk';\la,\spk)=-2\pi i e^2 \de(p\De k) \frac{e_i^{(\la)}(\spk)e_i^{*(\la')}(\spk')}{2 V \sqrt{k_0k_0'}} F(\De\spk),
\end{equation}
where
\begin{equation}
    F(\De\spk):=\lan e^{-i\De\spk\hat{\spx}}\ran.
\end{equation}
The latter quantity is nothing but the form-factor of the one-particle probability density
\begin{equation}
    N\rho(\spx):= N\sum_s\int\frac{d\spp d\spp'}{(2\pi)^3}e^{i(\spp-\spp')\spx} \rho_{ss}(\spp,\spp')=\int\frac{d\spk}{(2\pi)^3}e^{i\spk\spx}F(\spk).
\end{equation}
In the nonrelativistic limit, $\be\ll1$, or in the reference frame where the electron is at rest on average, under the fulfillment of the above assumptions and in the small quantum recoil limit, we obtain
\begin{equation}\label{perm_nonrel}
    \Phi(\la',\spk';\la,\spk)=-2\pi i \de(k_0-k_0')   \frac{e^2}{m} \frac{e_i^{(\la)}(\spk)e_i^{*(\la')}(\spk')}{2 V k_0} F(\De\spk),
\end{equation}
It is clear that the expressions obtained are also valid in the case $N=1$.

Comparing expression \eqref{perm_nonrel} with the amplitude of scattering of the electromagnetic waves by the susceptibility tensor  $\chi_{ij}(k_0;\spx)$ (see, e.g., formula (34) of the paper \cite{KazKor22}), we see that in the case we consider
\begin{equation}\label{suscept_plasm}
    \chi_{ij}(k_0;\spx)=-\frac{e^2N\rho(\spx)}{k_0^2m}\de_{ij},
\end{equation}
i.e., we reproduce the standard plasma susceptibility. In particular, such a susceptibility is inherent to the wave packet of a single electron. Choosing the normalization \eqref{d_norm}, the intensity of radiation \eqref{Compt_intens} in the comoving frame becomes
\begin{equation}\label{Compt_intens_expl}
    dI_{\la'}(\spk')=\Big|d_{\la'}(\spk')\sqrt{1+\mathcal{N}}-\frac{ie^2}{m}\sum_\la\int\frac{d\spk}{2(2\pi)^{2}k_0}\de(k_0'-k_0)F(\spk'-\spk) e^{*(\la')}_i(\spk') e^{(\la)}_i(\spk) d_\la(\spk)\Big|^2k'_0 d\spk'.
\end{equation}
It is not difficult to write the analogous formula in an arbitrary frame substituting expression \eqref{NrB1} in place of the second term under the modulus sing in \eqref{Compt_intens}. As we see, the contribution to the radiation intensity we consider describes the interference of the incident and reradiated photons and it is of the order $e^2$. In the case when this interference term is small, for example, for reflection of the photon from the electron wave packet (the inverse Compton scattering), it is necessary to take into account the corrections of the order $e^4$ to the radiation intensity, in particular, the standard contribution to the differential cross-section of the Compton process.

Just as for scattering of photons by dispersive media with susceptibility tensor possessing a certain symmetry, the Compton scattering by electron wave packet obeys the selection rules following from this symmetry. Let us examine how these selection rules look like for the symmetries considered in the previous section.
\begin{enumerate}[wide, labelwidth=!, labelindent=0pt]
  \item In this case,
\begin{equation}
     \Phi(\la',\spk';\la,\spk)\sim\de(k_3'-k_3).
\end{equation}
If the one-particle density matrix is invariant under arbitrary spatial translations, then
\begin{equation}
     \Phi(\la',\spk';\la,\spk)\sim\de(\spk'-\spk).
\end{equation}
The violation of translation symmetry of the one-particle density matrix on the scale $L$ results in smearing of the delta function $\de(k_3-k_3')$. The width of the peak becomes of the order $2\pi/L$.
  \item In this case,
\begin{equation}\label{Phi_periodic}
     \Phi(\la',\spk';\la,\spk)=\sum_{n=-\infty}^\infty \de(k_3'- k_3-qn) \Phi_n(\la',\spk';\la,\spk).
\end{equation}
By analogy with an ordinary periodic medium, one may say that the periodic wave packet imparts the projection of the momentum $\De k_3=qn$ to the scattered photon, i.e., the Bragg scattering is realized. It is interesting to note that this resonance scattering occurs even on the wave packet of a single particle.
  \item For the axially symmetric one-particle density matrix,
\begin{equation}
     \Phi(\la',\spk'_\vf;\la,\spk_\vf)=\Phi(\la',\spk';\la,\spk).
\end{equation}
Therefore, on scattering of twisted photons by such a state of the electrons, the projection of the total angular momentum onto the $z$ axis is conserved: $m'=m$.
  \item For the one-particle density matrix invariant under rotations $R_{\vf_r}$, we have
\begin{equation}
     \Phi(\la',\spk'_{\vf_r};\la,\spk_{\vf_r})=\Phi(\la',\spk';\la,\spk).
\end{equation}
The amplitude of scattering of twisted photons is proportional to
\begin{equation}
\begin{split}
    \int_{-\pi}^\pi\frac{d\vf'd\vf}{(2\pi)^2}e^{-im'\vf'+im\vf} \Phi(\la',\spk'_{\vf'};\la,\spk_\vf) &=\int_{-\pi}^\pi\frac{d\vf'd\vf}{(2\pi)^2}e^{i(m-m')\vf'}e^{im\vf} \Phi(\la',\spk'_{\vf'};\la,\spk_{\vf+\vf'})=\\
    &=\sum_{n=-\infty}^\infty\de_{m',m+nr}\int_{-\pi}^\pi\frac{d\vf}{2\pi} e^{im\vf} f_n(\la',\spk';\la,\spk_{\vf}),
\end{split}
\end{equation}
i.e., the selection rule, $m'=m+nr$, $n\in \mathbb{Z}$, holds. It is the same as for scattering of twisted photons by dispersive media possessing such a symmetry.
  \item In consequence of the helical symmetry, the representation \eqref{Phi_periodic} is valid and
\begin{equation}
    \Phi(\la',\spk'_{-\vf};\la,\spk_{-\vf})e^{i(k_3'-k_3)\vf/q}=\Phi(\la',\spk';\la,\spk).
\end{equation}
Hence
\begin{equation}
    \Phi_n(\la',\spk';\la,\spk)e^{in\vf}=\Phi_n(\la',\spk'_\vf;\la,\spk_\vf),
\end{equation}
and so
\begin{equation}
    \int_{-\pi}^\pi\frac{d\vf'd\vf}{(2\pi)^2}e^{-im'\vf'+im\vf} \Phi(\la',\spk'_{\vf'};\la,\spk_\vf)=\sum_{n=-\infty}^\infty \de_{m',m+n} \de(k_3'- k_3-qn) \int_{-\pi}^\pi\frac{d\vf}{2\pi} e^{im\vf}\Phi_n(\la',\spk';\la,\spk_\vf).
\end{equation}
Thus we deduce the selection rules for scattering by helical media \cite{KazKor22}.
  \item The combination of the symmetries 4 and 5 gives rise to the selection rules
\begin{equation}
    m'=m+rn,\qquad k'_3=k_3+qrn,\qquad n\in \mathbb{Z}.
\end{equation}
As for the intensity of radiation \eqref{Compt_intens_expl}, it represents the interference pattern of the incident and reradiated twisted photons.
\end{enumerate}

Notice that in order to observe the coherence effects in scattering of photons by electron wave packets, it is necessary that $k_0\gtrsim \s_e$. On the other hand, for $k_0\approx \s_e$,
\begin{equation}\label{chi_est}
    \chi_{ij}\approx-4\pi\al N\frac{\s_e}{m}\de_{ij},
\end{equation}
in the comoving frame. Keeping in mind that $\s_e\ll m$, the susceptibility of a wave packet of a single particle is rather small and it is challenging to detect the corresponding coherence effects. For example, for the photon wavelength $0.5$ $\mu$m, it follows from \eqref{chi_est} that
\begin{equation}
    \chi_{ij}\approx-7.1\times10^{-8}\de_{ij},
\end{equation}
when $N=1$. If the one-particle probability density $\rho(\spx)$ is of the order of $1/r_B^3$, where $r_B=1/(\al m)$ is the Bohr radius, then the plasma frequency corresponding to \eqref{suscept_plasm} for $N=1$ equals $\omega_p=2\sqrt{\pi} \al^2 m\approx 96.5$ eV.

\section{Conclusion}

Let us sum up the results. We investigated the processes of radiation of photons in QED with the initial states of the form \eqref{dens_matr_ini} up to the second order in the coupling constant $e$. We considered the $N$-particle, coherent, and thermal initial states of particles. In these cases, we obtained the general formulas for the intensity of radiation and the inclusive probability to record a photon. The background field method for nonperturbative treatment of a coherent initial state of photons with large amplitude was also discussed.

A special attention was paid to the processes where the wave packets of electrons radiate coherently, i.e., the electron wave packets can be regarded as some kind of a charged fluid in these processes. We found the three such processes in the second order of perturbation theory: stimulated radiation produced by an electron wave packet \cite{PanGov18,GovPan18,PanGov19,PanGov21,Talebi16}; coherent radiation from $N$ wave packets of particles arranged symmetrically, for example, as a bunch train \cite{MarcuseII,pra103}; reradiation by an electron wave packet in the Compton process. In the last case, the susceptibility tensor of a single electron wave packet in a vacuum was found.

We studied in detail stimulated transition radiation from the Dirac particle wave packet traversing a conducting plate irradiated by a laser wave. We obtained the explicit expression for the intensity of radiation and proved the selection rules for this radiation when the one-particle density matrix possesses certain symmetries. These selection rules appear to be the same as for transition radiation from beams of charged particles possessing the same symmetries, even in the case of stimulated transition radiation from a single electron. Therefore, this process can be used for mapping the profiles of electron wave packets employing the same techniques as for noninvasive diagnostics of beams of charged particles in transition and diffraction radiations \cite{Pbook,SukhKubPot17,SukhihDTHs}.

Then we investigated reradiation of photons by electron wave packets in a vacuum. We showed that this process is determined by the effective susceptibility tensor \eqref{Pi_eff_vac} appearing due to the presence of electrons in the initial state. We derived the explicit expression for this tensor and, as expected, it proved to be the same as for an electron plasma in the small recoil limit. However, what was not expected, this expression turns out to be valid even in the case of a single electron. Thus, we found the susceptibility tensor of an electron wave packet. We also obtained the explicit expression for the intensity of radiation produced in such a process and established the selection rules in the case when the one-particle density matrix possesses certain symmetries. These selection rules were shown to coincide with the selection rules for scattering of electromagnetic waves by a dispersive medium with the susceptibility tensor having the same symmetries. In particular, scattering of photons by a periodically modulated one-particle density matrix exhibits the Bragg resonances even in the case of scattering by a single electron.

As it has been already mentioned in Introduction, the formal cause for appearance of coherent radiation from a particle wave packet is the presence of contributions of through lines of the Feynman diagrams to the inclusive probability. These lines provide a free evolution of the wave packets of some particles participating in the process that, in turn, results in a coherent emission. Such a mechanism for coherent radiation from a particle wave packet is quite general and also appears in the processes different from those considered in the present paper. For example, in scattering of a muon by an electron, there is the contribution of the order $e^2$ to the inclusive probability to record a muon that represents an interference of the incident muon wave function with the correction to it due to scattering on the electromagnetic field produced by the Dirac current of the electron wave packet. In other words, this contribution is such as if the electron had moved freely and had not experienced a recoil.

\paragraph{Acknowledgments.}

This study was supported by the Tomsk State University Development Programme (Priority-2030).

%\newpage
\appendix
\section{Bargmann-Fock representation}\label{BF_represnt_Ap_A}

The states in the Bargmann-Fock representation,
\begin{equation}\label{BF_state}
    \Phi(\bar{a}):=\lan \bar{a}|\Phi\ran,\qquad\lan\bar{a}|a\ran=e^{\bar{a}a},
\end{equation}
satisfy the normalization condition
\begin{equation}\label{BF_norm_cond}
    \int D\bar{a}Dae^{-\bar{a}a}\bar{\Phi}(\bar{a})\Phi(\bar{a})=1.
\end{equation}
The functional integral is normalized such that
\begin{equation}\label{func_int_G}
\begin{split}
    \int D\bar{a}Da\exp\Big\{-\frac{1}{2}
    \left[
      \begin{array}{cc}
        a & \bar{a} \\
      \end{array}
    \right]
    B
    \left[
      \begin{array}{c}
        a \\
        \bar{a} \\
      \end{array}
    \right]
    +
     \left[
      \begin{array}{cc}
        a & \bar{a} \\
      \end{array}
    \right] F
    \Big\}&=
    \exp\big\{\tfrac{1}{2}F^T B^{-1}F\big\}
    \Big(\det\left[
      \begin{array}{cc}
        A_{21} & A_{22} \\
        A_{11} & A_{12} \\
      \end{array}
    \right]\Big)^{-1/2},\\
    \int D\bar{a}Da\exp\Big\{\frac{1}{2}
    \left[
      \begin{array}{cc}
        a & \bar{a} \\
      \end{array}
    \right]
    B
    \left[
      \begin{array}{c}
        a \\
        \bar{a} \\
      \end{array}
    \right]
    +
     \left[
      \begin{array}{cc}
        a & \bar{a} \\
      \end{array}
    \right] F
    \Big\}&=
    \exp\big\{\tfrac{1}{2}F^T B^{-1}F\big\}
    (\det B)^{1/2},
\end{split}
\end{equation}
where the first equality is for bosons ($\epsilon=1$), whereas the second one is for fermions ($\epsilon=-1$). The Grassmann parity of the column $F$ is equal to $(1-\epsilon)/2$. Besides,
\begin{equation}\label{B_matrix}
    B:=\left[
      \begin{array}{cc}
        A_{11} & A_{12} \\
        A_{21} & A_{22} \\
      \end{array}
    \right].
\end{equation}
The determinant on the right-hand side of the first line \eqref{func_int_G} is well-defined if $A_{11}$, $A_{22}$ are the Hilbert-Schmidt operators and $A_{12}-1$ and $A_{21}-1$ are the trace-class operators. As for fermions, the Gaussian functional integral is well-defined if the operators out of the diagonal of \eqref{B_matrix} are the Hilbert-Schmidt operators and the operators on the diagonal are of the form $1+(\text{trace-class})$. Furthermore, the Gaussian fermionic functional integral is also well-defined when one of the matrices,
\begin{equation}
    \pm B
    \left[
      \begin{array}{cc}
        0 & 1 \\
        -1 & 0 \\
      \end{array}
    \right],
\end{equation}
possesses the properties mentioned above (see for details \cite{BerezMSQ1.4}). In particular,
\begin{equation}\label{Gaussian_int}
    \int D\bar{a}Da e^{-\bar{a}a+\bar{a}\eta+\bar{\eta}a}=e^{\bar{\eta}\eta}.
\end{equation}
The trace of the operator in the Bargmann-Fock representation is given by
\begin{equation}\label{func_trace}
    \Sp\hat{A}=\int D\bar{a}Da e^{-a\bar{a}}\lan\bar{a}|\hat{A}|a\ran.
\end{equation}
Expressions \eqref{BF_state}, \eqref{BF_norm_cond}, \eqref{Gaussian_int}, and \eqref{func_trace} are valid for both statistics.

\section{Traces}\label{Traces_Ap_B}

In evaluating the one-particle and two-particle density matrices there appear the following expressions. For the $N$-particle Fock state \eqref{N_part_state}, the density matrix is
\begin{equation}
    \rho_{\al_N\cdots\al_1|\bar{\al}_1\cdots\bar{\al}_N}=\frac{|k|^2}{N!}\sum_{\s,\s'\in S_N}(-1)^{\e(\s)+\e(\s')}\vf^{\s'(N)}_{\al_N}\cdots \vf^{\s'(1)}_{\al_1}\bar{\vf}^{\s(1)}_{\bar{\al}_1}\cdots \bar{\vf}^{\s(N)}_{\bar{\al}_N}.
\end{equation}
Then, it is easy to show that the contractions \eqref{rho_1}, \eqref{rho_2} can be written in the form \eqref{rho_1_2_N}. As far as the thermal fermionic states are concerned,
\begin{equation}
    \lan\bar{a}|R_e|a\ran=\exp(\bar{a}e^{-\be\tilde{\e}}a)/Z,\qquad 1/Z=\det(1-n_F),
\end{equation}
we have
\begin{equation}
\begin{split}
    \Sp(R_e a^\dag_{\bar{\al}} a_\al)&=\int\frac{D\bar{a}DaD\bar{b}Db}{Z}\bar{b}_{\bar{\al}} a_\al\exp(\bar{a}a-\bar{b}b+\bar{b}a+\bar{a}e^{-\be\tilde{\e}}b)=\\
    &=-\frac{\de}{\de\bar{\eta}_\al}\frac{\overleftarrow{\de}}{\de\eta_{\bar{\al}}}\Big|_{\eta=\bar{\eta}=0} \int\frac{D\bar{a}DaD\bar{b}Db}{Z} \exp(\bar{a}a-\bar{b}b+\bar{\eta}a+\bar{b}\eta+\bar{b}a+\bar{a}e^{-\be\tilde{\e}}b).
\end{split}
\end{equation}
Using successively formulas \eqref{Gaussian_int} and \eqref{func_int_G}, we come to the first expression in \eqref{rho_therm_ferm}. In the same way,
\begin{equation}
\begin{split}
    \Sp(R_e a^\dag_{\bar{\al}_1}a^\dag_{\bar{\al}_2} a_{\al_2}a_{\al_1})&=\int\frac{D\bar{a}DaD\bar{b}Db}{Z} \bar{b}_{\bar{\al}_1} \bar{b}_{\bar{\al}_2} a_{\al_2}a_{\al_1}  \exp(\bar{a}a-\bar{b}b+\bar{b}a+\bar{a}e^{-\be\tilde{\e}}b)=\\
    &=\frac{\de}{\de\bar{\eta}_{\al_2}} \frac{\de}{\de\bar{\eta}_{\al_1}} \frac{\overleftarrow{\de}}{\de\eta_{\bar{\al}_2}} \frac{\overleftarrow{\de}}{\de\eta_{\bar{\al}_1}} \Big|_{\eta=\bar{\eta}=0} \int\frac{D\bar{a}DaD\bar{b}Db}{Z} \exp(\bar{a}a-\bar{b}b+\bar{\eta}a+\bar{b}\eta+\bar{b}a+\bar{a}e^{-\be\tilde{\e}}b),
\end{split}
\end{equation}
whence the second expression in \eqref{rho_therm_ferm} follows. The relations \eqref{rho_therm_bos} are proved along the same lines.

In order to find the inclusive probabilities, one needs to evaluate the following traces:
\begin{equation}
\begin{split}
    \Sp(P_{ph} c^\dag_{\bar{\ga}} R_{ph} c_{\ga})&=e^{-\bar{d}d}\frac{\de^2}{\de\bar{d}_\ga\de d_{\bar{\ga}}} \lan\bar{d}|P_{ph}|d\ran =e^{-\bar{d}d}\frac{\de^2}{\de\bar{d}_\ga\de d_{\bar{\ga}}} (e^{\bar{d}d}-e^{\bar{d}\tilde{D}d}),\\
    \Sp(P_{ph} c_{\ga}R_{ph} c^\dag_{\bar{\ga}})&=e^{-\bar{d}d}\bar{d}_{\bar{\ga}}d_\ga \lan\bar{d}|P_{ph}|d\ran =\bar{d}_{\bar{\ga}}d_\ga(1-e^{-\bar{d}Dd}),\\
    \Sp(P_{ph} c^\dag_{\bar{\ga}_1}R_{ph} c^\dag_{\bar{\ga}_2})&= e^{-\bar{d}d} \bar{d}_{\bar{\ga}_2}\frac{\de}{\de d_{\bar{\ga}_1}} \lan\bar{d}|P_{ph}|d\ran =e^{-\bar{d}d} \bar{d}_{\bar{\ga}_2}\frac{\de}{\de d_{\bar{\ga}_1}} (e^{\bar{d}d}-e^{\bar{d}\tilde{D}d}),\\
    \Sp(P_{ph} c_{\ga_1}R_{ph} c_{\ga_2})&= e^{-\bar{d}d} \frac{\de}{\de \bar{d}_{\ga_1}} \lan\bar{d}|P_{ph}|d\ran d_{\ga_1} =e^{-\bar{d}d} \frac{\de}{\de \bar{d}_{\ga_1}} (e^{\bar{d}d}-e^{\bar{d}\tilde{D}d}) d_{\ga_1},\\
    \Sp(P_{ph} c^\dag_{\bar{\ga}}c_{\ga}R_{ph})&=e^{-\bar{d}d}d_\ga\frac{\de}{\de d_{\bar{\ga}}}\lan\bar{d}|P_{ph}|d\ran= e^{-\bar{d}d}d_\ga\frac{\de}{\de d_{\bar{\ga}}} (e^{\bar{d}d}-e^{\bar{d}\tilde{D}d}),\\
    \Sp(P_{ph} c_{\ga_1}c_{\ga_2}R_{ph})&=e^{-\bar{d}d}d_{\ga_1}d_{\ga_2}\lan\bar{d}|P_{ph}|d\ran = d_{\ga_1}d_{\ga_2}(1-e^{-\bar{d}Dd}),\\
    \Sp(P_{ph} c^\dag_{\bar{\ga}_1}c^\dag_{\bar{\ga}_2}R_{ph})&= e^{-\bar{d}d} \frac{\de^2}{\de d_{\bar{\ga}_1} \de d_{\bar{\ga}_2}} \lan\bar{d}|P_{ph}|d\ran= e^{-\bar{d}d} \frac{\de^2}{\de d_{\bar{\ga}_1} \de d_{\bar{\ga}_2}} (e^{\bar{d}d}-e^{\bar{d}\tilde{D}d}),
\end{split}
\end{equation}
where
\begin{equation}
    R_{ph}=|d\ran\lan\bar{d}|e^{-\bar{d}d}.
\end{equation}

As for the thermal states of photons,
\begin{equation}
    \lan\bar{c}|R_{ph}|c\ran=\exp(\bar{c}e^{-\be\e}c)/Z,\qquad Z=\det(1+n_B),
\end{equation}
we obtain
\begin{equation}
\begin{split}
    \Sp(P_{ph} c^\dag_{\bar{\ga}} R_{ph} c_{\ga})&=\int \frac{D\bar{c}DcD\bar{d}Dd}{Z}
    \bar{c}_{\bar{\ga}}d_\ga\exp(-\bar{c}c-\bar{d}d+\bar{d}c +\bar{c}e^{-\be\e}d )(1-e^{-\bar{d}Dc})=\\
    &=\frac{\de^2}{\de\eta_{\bar{\ga}}\de\bar{\eta}_\ga}\Big|_{\eta=\bar{\eta}=0} \int \frac{D\bar{c}DcD\bar{d}Dd}{Z}
    \exp(-\bar{c}c-\bar{d}d+\bar{c}\eta+\bar{\eta}d+\bar{d}c +\bar{c}e^{-\be\e}d )(1-e^{-\bar{d}Dc}).
\end{split}
\end{equation}
Then applying sequentially formulas \eqref{Gaussian_int} and \eqref{func_int_G}, we arrive at the first expression in \eqref{tr_therm}. Analogously,
\begin{equation}
\begin{split}
    \Sp(P_{ph} c_{\ga}R_{ph} c^\dag_{\bar{\ga}})&=\int \frac{D\bar{c}DcD\bar{d}Dd}{Z}
    c_{\ga}\bar{d}_{\bar{\ga}} \exp(-\bar{c}c-\bar{d}d+\bar{d}c +\bar{c}e^{-\be\e}d )(1-e^{-\bar{d}Dc})=\\
    &=\frac{\de^2}{\de\eta_{\bar{\ga}}\de\bar{\eta}_\ga}\Big|_{\eta=\bar{\eta}=0} \int \frac{D\bar{c}DcD\bar{d}Dd}{Z}
    \exp(-\bar{c}c-\bar{d}d+\bar{d}\eta+\bar{\eta}c+\bar{d}c +\bar{c}e^{-\be\e}d )(1-e^{-\bar{d}Dc}),
\end{split}
\end{equation}
that gives the second expression in \eqref{tr_therm}. Furthermore,
\begin{equation}
\begin{split}
    \Sp(P_{ph} c^\dag_{\bar{\ga}}c_{\ga}R_{ph})=&\int \frac{D\bar{c}DcD\bar{b}DbD\bar{d}Dd}{Z}
    \bar{c}_{\bar{\ga}} b_{\ga} \exp(-\bar{c}c-\bar{b}b-\bar{d}d+\bar{c}b+\bar{d}c +\bar{b}e^{-\be\e}d )(1-e^{-\bar{d}Dc})=\\
    =&\frac{\de^2}{\de\eta_{\bar{\ga}}\de\bar{\eta}_\ga}\Big|_{\eta=\bar{\eta}=0} \int \frac{D\bar{c}DcD\bar{b}DbD\bar{d}Dd}{Z}
    \exp(-\bar{c}c-\bar{b}b-\bar{d}d+\bar{c}\eta+\bar{\eta}b+\bar{c}b+\bar{d}c +\bar{b}e^{-\be\e}d )\times\\
    &\times(1-e^{-\bar{d}Dc}),
\end{split}
\end{equation}
that results in formula \eqref{SpPccR_th}. The functional integral under the sign of variational derivatives with respect to $\eta$ and $\bar{\eta}$ is equal to the background contribution \eqref{background_therm} for $\eta=\bar{\eta}=0$.

%\slr

%\newpage

\end{document}